\documentclass[aps,preprint,floatfix,epsfig,tightenlines]{revtex4}
\usepackage{graphicx}

\begin{document}


\title{RICE Limits on the Diffuse Ultra-High Energy Neutrino Flux}

\author{I. Kravchenko}
\affiliation{Massachusetts Institute of Technology Laboratory for
Nuclear Science, Cambridge, MA  02139}
\author{C. Cooley}
\affiliation{Whitman College Dept. of Physics, Walla Walla, WA 99362}
\author{S. Hussain, D. Seckel, P. Wahrlich}
\affiliation{Department of Physics and
Astronomy and Bartol Research Institute, U. of Delaware, Newark DE 19716}
\author{J. Adams, S. Churchwell,
P. Harris, S. Seunarine}
\affiliation{Department of Physics and Astronomy,
Private Bag 4800, U. of Canterbury, Christchurch, New Zealand}
\author{A. Bean,
D. Besson,
S. Graham,
S. Holt,
D. Marfatia,
D. McKay,
J. Meyers,
J. Ralston$^*$,
R. Schiel,
H. Swift
}
\affiliation{University of Kansas Dept. of Physics and Astronomy, Lawrence KS
66045-2151}
\author{
J. Ledford,
K. Ratzlaff}
\affiliation{University of Kansas Instrumentation Design Laboratory, Lawrence KS 66045-2151}


\begin{abstract}
We present new limits on ultra-high energy
neutrino fluxes above
$10^{17}$ eV based on data
collected by the Radio Ice
Cherenkov Experiment (RICE) at the South Pole
from 1999-2005.
We discuss 
estimation of backgrounds, 
calibration and data analysis algorithms (both on-line and off-line), 
procedures used for
the dedicated neutrino search, and
refinements in our
Monte Carlo (MC) simulation, including
 recent {\it in situ} measurements of the complex
ice dielectric constant.
An enlarged data set and a more
detailed study of hadronic showers results in a
sensitivity improvement
of more than one order of magnitude compared to our previously
published results.
Examination of the full
RICE data set yields zero 
acceptable neutrino candidates, resulting in 95\% confidence-level
model dependent limits on the
flux $E_\nu^2d\phi/dE_\nu<10^{-6} {\rm GeV}/({\rm cm^2s~sr})$
in the energy range $10^{17}< E_\nu< 10^{20}$ eV. The new
RICE results rule out
the most intense flux model projections at
95\% confidence level.

$^*${\tt Contact for additional information.}
\end{abstract}

\maketitle


\newpage

\section{Introduction: UHE Neutrino Physics \label{s:intro}}
The motivation for producing a new map of the Universe, as viewed
through neutrino `eyes' is now well-established.
Ultra-high energy (``UHE''; E$>$1 PeV) neutrinos
point back to sources of high energy cosmic rays,
providing a direct picture 
of the source and the acceleration mechanism.
Detection of UHE neutrino fluxes simultaneous with 
gamma-ray bursts (GRB's)\cite{GRBs}
could provide essential information on the nature of
these extraordinarily luminous sources
and help resolve the question of whether GRB's are responsible for
the bulk of the UHE cosmic ray particle flux incident at Earth.
In the $10^{18}-10^{20}$ eV energy regime,
``GZK'' neutrinos may distinguish between source evolution models for UHE
cosmic rays\cite{SeckelS05}. At even higher energies, ``Z-burst" neutrino
models\cite{Weiler} have been proposed to 
explain the $\ge$$10^{20}$~eV cosmic ray flux claimed
by the AGASA experiment\cite{AGASA98}; neutrinos may also
identify more
exotic sources such as topological defects\cite{Berezinsky}.
In the realm of particle physics, detection of UHE neutrinos
from cosmological distances, if
accompanied by flavor identification, may permit measurement
of neutrino oscillation parameters
over a wide range of
$\Delta m^2$\cite{Francis-David-mixing-sensitivity,Pakvasa03} or
observation of $\nu_\tau$ via ``double-bang'' signatures\cite{doublebang}.
Additionally,
the angular distribution of upward-going neutrino events
could be used to measure weak
cross-sections at energies unreachable by man-made 
accelerators\cite{sigma}.
Alternately, if the high energy weak cross-sections are known,
they can be used to test Earth composition models
along an arbitrary cross-section, so-called
`neutrino tomography'\cite{tomography}.

Several recent projects
(AMANDA\cite{AMANDA1}, IceCube\cite{IceCube},
NESTOR\cite{NESTOR}, NEMO\cite{NEMO},
Lake Baikal\cite{BAIKAL},
ANTARES\cite{Antares}, e.g.)
have demonstrated photomultiplier-tube based detection of 
high energy cosmic ray
muon neutrinos, by observing
the optical Cherenkov cone which results from muons
produced in
$\nu_\mu$ weak current interactions. 
To detect neutrinos at UHE, it is more effective to exploit coherence of radio Cherenkov emissions. 
The amplitude of radio frequency signals emitted by neutrino-generated electromagnetic showers in the MHz-GHz regime increases nearly linearly with energy, making radio detection the most efficient scheme presently known 
at ultra-high energies.  

RICE employs this strategy to search for neutrino interactions occurring in cold polar ice, which has exceptional transmission properties favorable for radio detection.  Here we extend our previous results \cite{astroph-cal,astroph-results}, which were based on shorter running times and less advanced detector response modeling.  We report new limits based on additional data and further consideration of systematic errors.

\section{The RICE Detector \label{s:detect}}
The status of the current array
deployment is summarized in Table \ref{tab:Rxlocations};
further details on detector geometry, deployment and calibration procedures
are presented elsewhere\cite{astroph-cal}.
The Martin A. Pomerantz Observatory (MAPO) building houses hardware for several experiments, including the RICE and AMANDA surface electronics, and is centered at (x$\sim$40m, y$\sim -30$m) on the surface. The AMANDA array is located approximately 600 m (AMANDA-A) to 2400 m (AMANDA-B) below the RICE array in the ice; the South Pole Air Shower Experiment (SPASE) is located on the surface at (x$\sim$-450m, y$\sim$0m). The planned IceCube experiment will circumscribe the existing AMANDA experiment, with a hexagonal footprint of radius $\sim$500 m and centered approximately 200 m northwest of the origin.
The coordinate system conforms to the convention used by the 
AMANDA experiment: 
grid North is defined by the Greenwich Meridian. 
\begin{table}[hptb]\begin{tabular}{c|c|c|c}Channel (Hole) & x- (m) & y- (m) & z- (m) \\ 
\hline 0 (A11)& 4.8 &  102.8 &  -166 \\ 1 (A6) & -56.3 &  34.2 &  -213  \\ 2 (A13) & -32.1 &  77.4 &  -176 \\ 3 (A12) & -61.4 &  85.3 &  -103 \\ 4 (A6) & -56.3 & 34.2 & -152 \\ 5 (A7) & 47.7 & 33.8 & -166 \\ 6 (B2) & 78.0 & 13.8 & -170 \\ 7 (B3) & 64.1 & -18.3 & -171 \\ 8 (B1) & 43.9 & 7.3 & -171 \\ 9 (B3) & 64.1 & -18.3 & -120 \\ 10 (B1) & 43.9 & 7.3 & -120 \\ 11 (B4) & 67.5 & -39.5 & -168 \\ 12 (A18) & 66.3 & 74.7 & -110 \\ 13 (A15) & -95.1 & -38.3 & -105 \\ 14 (A16) & -46.7 & -86.6 & -105 \\ 15 (A19) & 95.2 & 12.7 & -347 \\ 19 (A15) & -95.1 & -38.3 & -135 \\ \hline \end{tabular} \caption{\it \label{tab:Rxlocations}Location of RICE radio receivers for the bulk of the data relevant to this paper (in Jan. 2004, channel 11 was moved to just below the surface, at a depth of approximately 2 m). We have adopted the coordinate system convention used by the AMANDA collaboration. ``A'' holes correspond to holes drilled for AMANDA; holes B2 and B4, drilled for RICE in 1998, have been re-opened in subsequent seasons for radioglaciological measurements.} \end{table}

\subsection{Radio Cherenkov Detection \label{ss:RCD}}
Long-wavelength (radio-wave) detection of electromagnetic showers in
dense media relies on two fundamental pillars - 
long attenuation lengths of order 1 km in cold polar ice, and 
 coherence  
extending up to 1 GHz for radio Cherenkov emission from the net
charge developing in showers, as recently
verified using data taken in an electron testbeam\cite{slac-testbeam,slac-salt04}. 
Ultra-high energy showers in dense targets contain  
roughly one excess electron per four GeV of shower energy, leading to a
rapid growth of sensitivity with increasing energy. 
The ANITA\cite{ANITA}, GLUE\cite{GLUE},
FORTE\cite{FORTE},
RAMAND\cite{RAMAND}, 
RICE\cite{astroph-results,astroph-cal,RICEnZ},
and SALSA\cite{SALSA} projects all 
seek radiowave neutrino
detection in dense media. Other efforts\cite{JonRosner,LOFAR,CODALEMA} 
are
directed toward radiowave detection of atmospheric cascades. 
Radio detection schemes have also received
considerable attention
as probes of monopoles\cite{WickWeiler}, 
TeV-scale gravity\cite{Feng1,Feng2,Shahid03,Shahid05,Pankaj02}, 
and tau-neutrinos\cite{Hallsie-2003}.

Discussions of the Askaryan effect\cite{Askaryan} upon
which the radiowave detection technique is founded, its experimental
verification in a testbeam environment\cite{slac-testbeam,slac-salt04}, 
calculations
of the expected radio-frequency signal from a purely electromagnetic
shower\cite{ZHS,Alvarez-papers,SoebPRD,addendum,RomanJohn},
as well as hadronic showers\cite{Shahid-hadronic}, and modifications
due to the LPM effect\cite{LPM,spencer04} can be found in the literature.
RICE uses its own Monte Carlo-based procedure based on 
GEANT4-generated showers and charge-by-charge superposition of Cherenkov 
radio emissions\cite{SoebPRD,addendum} to estimate the signal
strength.
Several simulations\cite{ZHS,SoebPRD,Butkevich} now give consistent
estimates for the expected charge excess, as well as the 
electric
field signal.
For frequencies and geometries relevant to RICE,
we estimate the uncertainty in the signal field strength,
for $f<$1 GHz to be $\lesssim$10\%, although errors associated with
the LPM effect may be somewhat larger. 

Sections \ref{s:CDS}-\ref{s:SU} below review calibration of the array, event 
reconstruction, determination of the effective volume, data acquired  
and potential events. Section \ref{s:NFLR} presents the new limits, and
sections \ref{sect:Summary} and \ref{s:FWFP}
present a summary and outlook. Appendices present more details on the 
calculation of our upper limits, as well as a procedure for deriving 
the sensitivity of RICE to any arbitrary flux model.

\message{ADD ALL RUNTIMES, ETC.}
\section{Current Data Set \label{s:CDS}}
The data taken thus far with the RICE array are summarized in
Table \ref{tab:datasum}. 
\begin{table}
\begin{center}
\begin{tabular}{c|ccccccc|c}
	& 1999 & 2000 & 2001 & 2002 & 2003 & 2004 & 2005 & {\bf Total} \\ \hline
Total RunTime ($10^6$ s) & 0.18 & 22.3 & 4.6  &  19.9 & 24.5 & 11.6 &  15.1 & 98.2\\
Total LiveTime ($10^6$ s) & 0.10 & 15.7 & 3.3   & 13.6 & 17.1 & 9.4 & 14.9 & 74.1\\
DeadTime (303 ON) ($10^6$ s)   & 0.03 &  3.7 & 1.0        & 4.1    & 5.6 & 1.1 & 0.0 & 15.5 \\
$\geq$4-hit General Triggers ($\times 10^4$)   & 0.26 & 30.6 & 6.0    & 16.9 & 13.8 & 9.4 & 26.5 & 103.5\\
Unbiased Triggers ($\times 10^4$)    & & 3.3 & 1.3   & 3.5   & 4.4  & 2.5 & 4.0 & 19.0 \\
AMANDA-coincident Triggers ($\times 10^4$)   & 0.064 & 1.9 & 2.4 & 0.016 & 0.056 & 0.075 & 0.002 
& 4.51 \\
SPASE-coincident Triggers ($\times 10^4$)   & & 0.48 & 0.003         & 0.47           & 0.021    & 0.001 & 0.067 & 1.04 \\
Veto Triggers ($\times 10^4$)   & 1.2 & 11182.8 & 317.4   & 12973.9
& 3153.9 &  142.5 & 471.0 & 28242.7 \\ \hline 
\end{tabular}
\caption{\it Summary of RICE-II data taken through Aug. 15, 2005.
Time (303ON) is the total time (in seconds) that the 
303 MHz
South Pole Station satellite
uplink to the LES communications satellite was active and prohibited
data-taking.
``4-hit Triggers'' refer to all events for which there are
at least
four RICE antennas registering voltages
exceeding a pre-set discriminator threshold 
in a coincidence time comparable to the light
transit time across the array ($1.25\mu$s); ``Unbiased
Triggers'' correspond to the
total number of events taken at pre-specified intervals
and are intended to capture
background conditions within the array; 
``AMANDA-coincident Triggers'' correspond to 
events for which there is at 
least one RICE antenna hit within
$1.25\mu$s of an AMANDA high-multiplicity 
Optical Module trigger (this trigger
has been intermittently disconnected during the course of the experiment when
the trigger rate became prohibitively large);
``SPASE-coincident Triggers'' correspond to events
for which there is at least one RICE antenna hit within 
$1.25\mu$s of a SPASE high-multiplicity surface scintillator trigger
and have been used in a search for coincident air shower detections;
``Veto Triggers'' are events tagged online by a fast ($\sim$10 ms/event)
software algorithm as consistent with 
having a surface origin. Full DAQ readout of
such events is heavily pre-scaled,
typically by a factor
of 10000, to mitigate the compromise in livetime incurred in writing data to disk. Data-taking problems in 2001 resulted in a 
reduced accumulated livetime for that year.}
\label{tab:datasum}
\end{center}
\end{table}
Over a typical 24-hour period, roughly 1000 data
event triggers currently
pass a fast online hardware surface-background veto
($\sim$1$\mu$s/event) and an online software surface-background
veto ($\sim$10 ms/event). To these data we have applied a sequence 
of offline cuts to remove background, as detailed later in this document.
We determine the efficiency of our
event selection criteria using simulations of showers, both
electromagnetic and hadronic, resulting from neutrino
collisions, superimposed on environment characterization drawn 
from data itself (unbiased events). 

\section{Backgrounds\label{s:bkgnds}}
We generally distinguish the different backgrounds to
the neutrino search according to the following
criteria:
a) vertex location of reconstructed
source, b) waveform characteristics (time-over-threshold, e.g.) of
hit channels, c) goodness-of-fit to a 
well-constrained single vertex as evidenced by
timing residual characteristics (discussed in more detail below),
 d) RF conditions during data-taking,
e) Fourier spectrum of hit channels, f) cleanliness of hits
(e.g., presence of multiple
 pulses in an 8.192 microsecond waveform capture),
g) multiplicity of receiver antennas registering hits for a particular event,
h) time-since-last-trigger ($\delta t_{ij}\equiv t_i-t_j$, 
where $t_i$ is the time of the
$i^{th}$ trigger and $t_j$ is the time of the next trigger. In high-background,
low-livetime instances, we expect 
$\delta t_{ij}\to \delta t_{min}$, 
where
$\delta t_{min}$ is the $\sim$10 s/event readout time of the DAQ. In
low-background, high-livetime instances, we expect
$\delta t_{ij}\to \delta t_{max}$, where $\delta t_{max}$ is the
ten-minute interval between successive unbiased triggers), and i) 
trigger type fractions.
We can coarsely
characterize three general 
classes of backgrounds according to the
above scheme, as follows. 

1) Continuous wave backgrounds (CW) are expected to have
a) a long time-over-threshold
for channels with amplitudes well above the
discriminator threshold, b) large timing residuals (since the hit times will
not be correlated with a single source), c) small values of
$\delta t_{ij}$ for the case where the discriminator threshold is
far below the CW amplitude, 
d) backgrounds occurring in all trigger types, since 
the background will be present in ``unbiased'' 
forced-trigger events as well as ``general''
4-hit
events, e) a Fourier spectrum dominated by one frequency (plus overtones),
f) a hit multiplicity which is on 
average roughly
constant, and determined by the number of channels which
exceed threshold when their noise voltage is added to the underlying
CW voltage. Such backgrounds may cluster in time or show
a diurnal periodicity and are generally easily recognized on-line.

2) True thermal noise backgrounds should have a) vertex locations which are 
spatially
distributed as Gaussians  
centered at the centroid of the array (x=0, y=0, z=--120 m), as demonstrated by
Monte Carlo (by simulating
four hits at random times 
within a 1.2$\mu$s discriminator window,
roughly corresponding to the light transit
time across the array; see Figs. \ref{fig:dataVnoise.ps} and
\ref{fig:aug2000-vx-vy-therm}), 
b) very small time-over-thresholds with signal shapes which
are largely indistinguishable from a true neutrino-induced
signal -- for this reason, thermal noise events satisfying all
other kinematic selection criteria are
also likely to pass a visual
event hand-scan, c) large timing
residuals (Fig. \ref{fig:thermtrigs}), d) successive
trigger time difference characteristics which depend in a statistically
predictable way on the ratio of discriminator thresholds to rms thermal noise
voltages, 
e) a ratio of general/unbiased triggers which, in principle, can be
statistically derived from the thermal noise distribution observed in 
unbiased events, f) a Fourier spectrum dominated
by the bandwidth of the various components of a RICE receiver circuit, 
g) no double pulse characteristics, 
h) no correlation with date or time. 
Unbiased events are expected to be representative of the
experimental thermal noise 'floor'.
In practice,
examination of a large number of
unbiased events show non-Gaussian tails in a large
fraction of the voltage distributions, 
indicating that there are non-thermal backgrounds present in many of
these events.

3) ``Loud''
transients are observed to constitute the
dominant background.
We sub-divide possible transient
sources into two categories: those sources which originate within the ice
itself, primarily due to AMANDA and/or
IceCube phototube electronics, and those
sources which originate on, or
above the surface. 
Without filtering, the extremely large amplitude radiofrequency
transients generated by each AMANDA/IceCube phototube (typically,
two km distant and triggering
at $\sim 10^2$ Hz) would result in a prohibitively large background.
After our initial deployment of three
test antennas in 1996-97, highpass 
($>$250 MHz) filters were inserted
to suppress these backgrounds, leaving 
more sporadic anthropogenic
surface-generated noise as the dominant transient background.
Such triggers
are characterized by: 
a) typically, large time-over-thresholds, 
b) $\delta t_{ij}$ distributions which reflect saturation
of the DAQ, or 
show structure if the source is periodic, c) Fourier spectra which 
are likely to show non-thermal structure.

\subsection{Vertex
Suppression of Transient Anthropogenic Backgrounds \label{s:TAB}}

Vertex distributions
give perhaps the most direct characterization 
of surface-generated (z$\sim$0) vs.
non-surface (and therefore, candidates for more interesting processes) events.
Consistency between various source reconstruction algorithms gives 
confidence that the true source
has been located.
Due to ray tracing
effects, it is difficult to identify surface sources at large
polar angles, which increasingly fold into the region around
the critical angle.
We implement both a ``grid''-based
vertex search algorithm, as well as an
analytic, 4-hit vertex reconstruction algorithm, as 
detailed previously\cite{astroph-cal}.
Once a vertex has been found,
one discriminant of ``well-reconstructed'' vs. 
``poorly-reconstructed'' sources is provided by calculating the average
time residual per hit. This is done by: 1) identifying a putative vertex
for the event, 2) calculating the expected recorded hit time for each
channel assuming that vertex, after
taking into account ice-propagation
time plus cable delays plus electronics propagation delays at the surface,
3) calculating the difference between the expected time and the 
actual, measured time, for that reconstructed vertex.
That time difference is defined as the ``time
residual'' (as defined for fits to the helical
trajectory expected for a charged track traversing a multi-layer drift 
chamber) for that particular channel. 
This parameter, 
when minimized over all channels, defines the reconstructed vertex using
the grid-based algorithm. 
By constrast, the per-channel 
spatial residual is defined as the distance between
the reconstructed vertex when a given channel is included in
vertex reconstruction vs. excluded from vertex reconstruction,
and is therefore only defined for events with multiplicity greater than
or equal to five.
 
\begin{figure}
\centerline{\includegraphics[width=9cm]{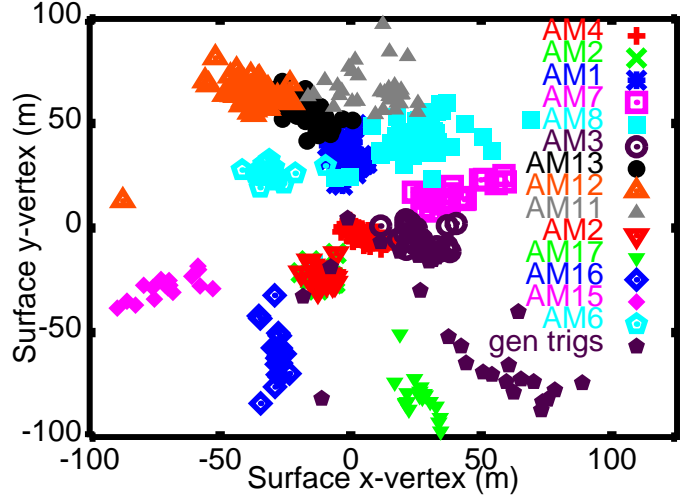}}
\caption{\it Reconstructed xy-vertices for transmitter data taken
with transmitter located above the indicated AMANDA hole, compared with
a small sample of general triggers. 
True source surface coordinates are
as indicated in Table I.}
\label{fig:AM-gen-trigs.eps}
\end{figure}

To estimate our ability to correctly reconstruct source vertices,
transmitters were placed at various points on the surface, as well
as at various depths within the ice, and the reconstructed 
vertex location compared with the known vertex location.
Figure
\ref{fig:AM-gen-trigs.eps} was obtained by broadcasting down to the
RICE array from an elevated (z$\sim$+3 m) dipole
positioned atop the surveyed AMANDA holes
(Table \ref{tab:Rxlocations}), and then reconstructing
the source location using our standard timing methods.
Our surface source reconstruction
resolution within this solid angle is of order 10 m, with poorer longitudinal 
(vs. lateral) resolution. 

\begin{figure}
\centerline{\includegraphics[width=9cm]{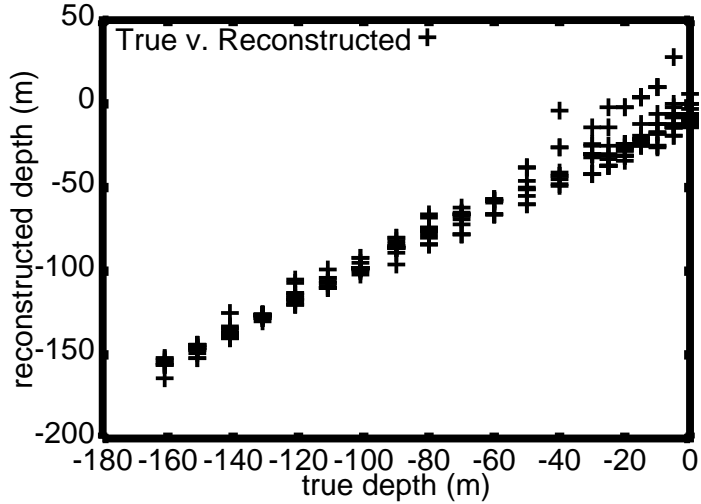}}
\caption{\it Reconstruction of z-coordinate of a transmitter as it is lowered
into RICE hole B4. For analytic
vertex reconstruction, we assume a constant value of index-of-refraction n=1.}
\label{fig:B4fits0}
\end{figure}

To estimate our ability to reconstruct sources at shallow depths
(e.g., buried active electronics around South Pole Station), a
dipole transmitter was pulsed as it was lowered into a hole in the vicinity
of the RICE array (hole B4). Figure \ref{fig:B4fits0} shows the result of
this exercise, and also illustrates the
expected degradation in resolution as z$\to$0. In our subsequent
neutrino analysis we require that the reconstructed source depth be greater
than 200 m.

\subsection{Transient and CW Diurnal Backgrounds \label{ss:transient}}
\begin{figure}
\includegraphics[width=9cm]{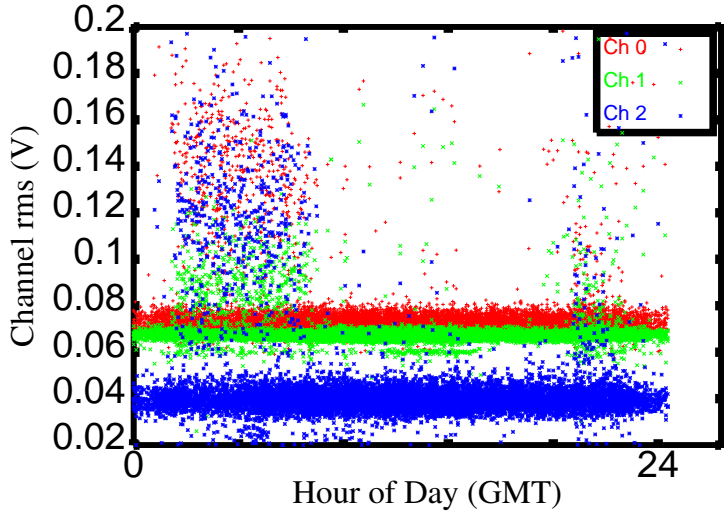}
\caption{\it Recorded rms voltage, for three channels, as a function of time
of day, for data taken between 1/4/04 and 6/4/04.}
\label{fig:rms-log-4jan04-5jan04}
\end{figure}

We may expect that anthropogenic backgrounds might be periodic with
a 24-hour timescale. Figure \ref{fig:rms-log-4jan04-5jan04} 
shows the measured
mean rms voltage in three channels as a function of time of day.
\message{HOW DOES THIS FIGURE GO UP TO 25 HRS?}
There is clear evidence for periodic backgrounds, such as the station
satellite uplink during those times when communications satellites are
above the horizon, although not all background sources have yet been
fully identified\cite{Corinne-thesis}.

\subsection{Thermal Noise backgrounds \label{ss:TNB}}
\begin{figure}
\centerline{\includegraphics[width=9cm]{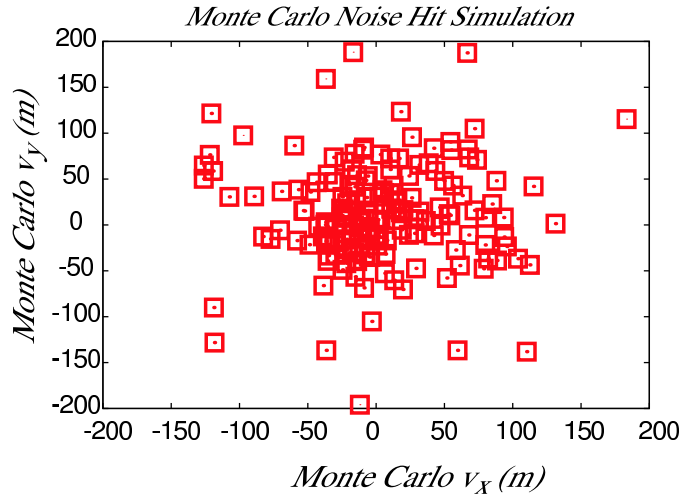}}
\caption{\it Simulation of x-coordinate of reconstructed vertex ($v_x$)
vs. y-coordinate ($v_y$) of reconstructed vertex for ``noise'' hits generated
in Monte Carlo simulations.}
\label{fig:dataVnoise.ps}
\end{figure}

\begin{figure}
\centerline{\includegraphics[width=9cm]{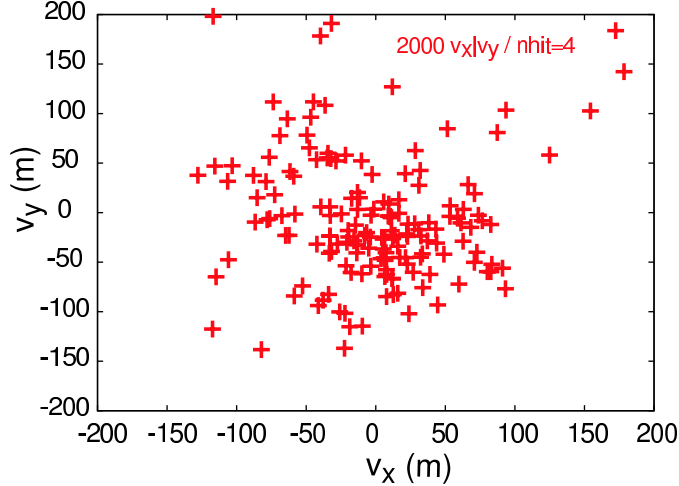}}
\caption{\it August 2000 data, $v_x$ vs. $v_y$ vertex distribution, for events
having low hit-multiplicity ($N_{hit}$=4) and large time-between-successive-triggers. Events are required to pass initial time-over-threshold requirements.}
\label{fig:aug2000-vx-vy-therm}
\end{figure}

When anthropogenic backgrounds are low
and the experiment is operating close to the thermal limit, the
reconstructed vertex distribution for thermal noise
events is expected to peak close to the
center of the array, with a width given by the light transit distance
across the 1.25 $\mu$s
coincidence window defined by the RICE general event trigger.
Monte Carlo expectations for the vertex distributions reconstructed
from such ``thermal events'' are shown in 
Figure \ref{fig:dataVnoise.ps}.
By comparison, the vertex distributions
for data during a time when the detector was dominated
by thermal noise hits (August 2000, as determined by the preponderance
of unbiased triggers in those data) 
are shown in 
Fig. \ref{fig:aug2000-vx-vy-therm}.
During the winter months, when station noise is typically lowest, 
approximately 50\% of our backgrounds are thermal noise backgrounds.
During the austral summer months, when human activity at
South Pole Station is largest, this fraction decreases to less than
10\%.

\message{TBD:QUANTIFY THE
THERMAL-LIKE FRACTION OF EVENTS IN OUR DATA, AND
PRESENT A PLOT, ALSO GIVE A FRACTION FOR
SURFACE EVENTS, AS WELL AS UNEXPLAINED HIGH NXTOT EVENTS.}

\subsection{Showers from Atmospheric Muons \label{ss:muon}}

High energy muons produced in cosmic ray interactions may penetrate into 
the ice and suffer catastrophic $dE/dX$ bremsstrahlung or photonuclear 
interactions. These interactions produce in-ice showers potentially 
visible to RICE. The magnitude of this background can be estimated from 
Figs. 3 and 8 of Ref. \cite{Thunman}.  Of interest to RICE,  the vertical 
muon flux above 100 PeV is $E\Phi(E)\sim 2.5 \times 10^{-3} 
({\rm 100~PeV}/E_\mu)^3$ km$^{-2}$sr$^{-1}$yr$^{-1}$. 
The rate increases with 
zenith angle, roughly as $1/\cos \theta_z$. A typical UHE muon is 
expected to bremsstrahlung one or two photons with $\sim$10\% of $E_\mu$ 
over a 1 km track length in ice; at energies above $\sim 200$ PeV 
photonuclear interactions increase the shower rate. Combining these 
factors, we estimate that when averaged over the full sky, the integrated 
rate for shower production with $E_s > 100$~PeV is a few times 10$^{-5}$ 
km$^{-3}$sr$^{-1}$yr$^{-1}$.

As we show below, the RICE integrated exposure for 100 PeV 
showers is of order 1 km$^{3}$sr$^{1}$yr$^{1}$. Taken together these 
factors lead us to expect $<$10$^{-4}$ muon-induced events in 
our data sample. Decreasing the energy reduces the experimental
sensitivity, whereas 
increasing the energy reduces the muon flux; at $\sim$100 PeV the RICE
response for this signal is maximal.

We note that the estimated rate depends on an uncertain extrapolation of 
charm hadroproduction cross sections from low energy, and that current 
experimental bounds allow a flux up to 100 times higher than shown in 
Ref. \cite{Thunman}. Still, we expect less than 0.01 events from this 
source even in the most optimistic scenarios.

\subsection{Atmospheric Neutrinos \label{ss:atmu}}

Above 1 PeV, the atmospheric neutrino background is also dominated by 
charm production\cite{Thunman}. The $\nu_\mu$ and $\nu_e$ fluxes are 
slightly larger in magnitude than the atmospheric muon flux. Summing over 
flavor and averaging over the full sky, we expect a total neutrino flux 
above 100 PeV of $E\Phi(E) \sim 2 \times 10^{-20} ({\rm 100~PeV}/E)^3$ 
cm$^{-2}$sr$^{-1}$s$^{-1}$. RICE Monte Carlo studies indicate a 
sensitivity of about $\sim 10^{14}$ cm$^{2}$sr$^{1}$s$^{1}$ at 100 PeV. 
The mismatch between flux and sensitivity
($10^6$) is slightly greater than for atmospheric muons. 
Although a larger fraction of the neutrino energy will be converted to 
shower energy, only $\sim 10^{-3}$ of the neutrinos interact while passing 
through the RICE sensitive volume. Even with enhanced charm production we 
expect no atmospheric neutrino events in the RICE data set. We also note 
that the RICE effective volume is still small below 10 PeV, so we have 
very little sensitivity to the Glashow resonance at $E_{\nu_e} \sim 
6.4$~PeV.

The small fluxes of atmospheric muons and neutrinos above 100 PeV allow
RICE to circumvent the primary neutrino backgrounds confronting the optical 
Cherenkov experiments. However, these small rates also deprive RICE of 
an obvious calibration `beam'.

\subsection{Flaring Solar RF Backgrounds \label{ss:flares}}
Radio frequency noise associated with solar activity has been
the subject of extensive investigation. Auroral discharges have
been continuously monitored in Antarctica in the tens of MHz frequency range,
over the last decade. In 2003, there were high-intensity solar 
flares recorded between Oct. 19, 2003 and Nov. 4, 2003; typically, these
result in electrical disturbances on Earth about 24-48 hours 
later\cite{000714}.
We have searched for correlations during this time period with
high data-taking rates as registered by RICE.
We observe no obvious evidence for correlation of our trigger rates
with solar flare activity\cite{Corinne-thesis}.

\subsection{Air Shower Backgrounds \label{ss:air-shower-backgrounds}}

Complementing the production of UHE muons and neutrinos discussed above, 
there are several possible radio signals associated directly with cosmic 
ray air showers. These include the production of geo-synchrotron 
radiation in the atmosphere, as well as transition and Cherenkov signals 
produced as the shower impacts and evolves into the ice. These three 
mechanisms all require coherent radiation from all or part of the shower. 
In all three cases, the transverse profile of the shower dictates a 
fundamental frequency response, whereas for the geo-synchrotron and 
Cherenkov signals the shower/observer geometry must also be favorable to 
have coherent emission from the full longitudinal development of the 
shower.

Coherent production of synchrotron radiation in the geomagnetic field 
has recently been observed by the LOPES\cite{Falcke05} and 
CODALEMA\cite{Ardouin05} collaborations. This signal is most interesting 
below 100 MHz\cite{HuegeF05}, and falls off rapidly in the RICE bandpass. 
We have not studied this mechanism in detail, but note that the frequency 
response is ultimately related to the geometry of the air shower -- the 
signal rolls over at $f\sim R/r_M^2$ where R$\sim$2-3 km is the 
height of shower max and $r_M \sim 100$~m is the Moliere radius for the 
shower.

Transition radiation results when the shower impacts the ice\cite{gazazian}. 
In this 
case, R$\sim$200~m for RICE, f$\sim$200~MHz, and the region for 
coherent emission is a disk of order 10~m radius. Only a fraction of the 
excess shower charge is contained within that distance of the shower axis. 
Further, transition radiation is forward peaked, so illumination of more 
than one antenna is rather unlikely. We have not seriously modeled  
transition radiation from air shower impacts as a background for RICE.

The most interesting signal for RICE is the Askaryan pulse produced when 
the air shower core hits the ice. At RICE frequencies, the Askaryan pulse 
must originate from a transverse dimension comparable to that for a 
shower initiated in-ice, a few tens of cm at most. This length scale is 
compatible with the core of the shower where the highest energy particles 
reside. Particles have their last interactions of order 1 km above the 
ice, so the required relativistic-$\gamma$ factor is of order $10^4$, 
corresponding to particle energies $\sim 10$~GeV for $e^-$, $e^+$ and 
bremstrahlung $\gamma$'s. 

Accordingly, we generated vertical proton showers using the default South 
Pole configuration of ARIES\cite{AIRES}, keeping track of all particles 
reaching the ground with energies greater than 10 GeV.  We find that for 
PeV protons, typically $\sim 3\%$ of the primary energy impacts within 30 
cm of the position of the primary axis. This energy is available for 
producing an Askaryan pulse. We expect that as the primary energy 
increases a larger fraction of the energy remains in the core, but as we 
consider less vertical showers, the core will weaken. A rough estimate is 
that for a pure proton composition the rate of surface showers with 
energy above 1 EeV is comparable to the rate of 10 EeV primaries 
averaged over 1 sr, or roughly 0.5 event per km$^2$ yr. 

We are in 
the process of enhancing the RICE Monte Carlo to model the modified 
Askaryan pulses which develop in the low density snow/firn at the 
surface, and are making the necessary modifications for ray tracing and 
antenna response in this geometry. Even if the signal is significant, 
such events may not pass the analysis chain designed largely to eliminate 
surface noise of anthropogenic origin. We have looked through one year (2002)
of data\cite{Wahrlich05}, but no clear coincidences with the SPASE array 
in our SPASE-trigger sample
have been observed.

\message{FOR AIR SHOWERS: A)
 CAN WE SKIM EVENTS WHICH ARE AT THE EDGE OF THE C-CONE
CAUSTIC, POINTING UP, AND INCONSISTENT WITH STUFF DIRECTLY OVERHEAD OR
STUFF POINTING TOWARDS SPASE OR THE DOME? B) IS THERE A MAGNIFICATION EFFECT
WHEREBY ALL SIGNAL CLOSE TO THE SURFACE IS FUNNELED INTO RICE??

more thoughts on air showers - all signal outside 45-degree angle folds into
45-degree angle. How much signal is opposite MAPO? Alternately, how many
times do we see signal sweeping in from channel 0?!

NEED TO LOOK AT HIGH-MULTIPLICITY SPASE TRIGGERS (TRIGCODE=3)! WHY SHOULD
THERE BE ANY HIGH-MULTIPLICITY EVENTS DUE TO BACKGROUND IF WE REQUIRE ONLY
A 1.2 MICROSECOND WINDOW FOR A 1-HIT COINCIDENCE. ALSO NEED TO KNOW THE
RAW SPASE TRIGGER RATE FOR THEIR 30-FOLD COINCIDENCE TO FIRE!

SPASE.gnu - why do SPASE triggers have high-multiplicity???

INPUT NEW TEXT FROM JENNI/SURUJ!}


\section{Detector Calibration and Modeling Details \label{s:DCM}}
\subsection{Antenna Response \label{ss:AR}}
\begin{figure}
\centerline{\includegraphics[width=9cm]{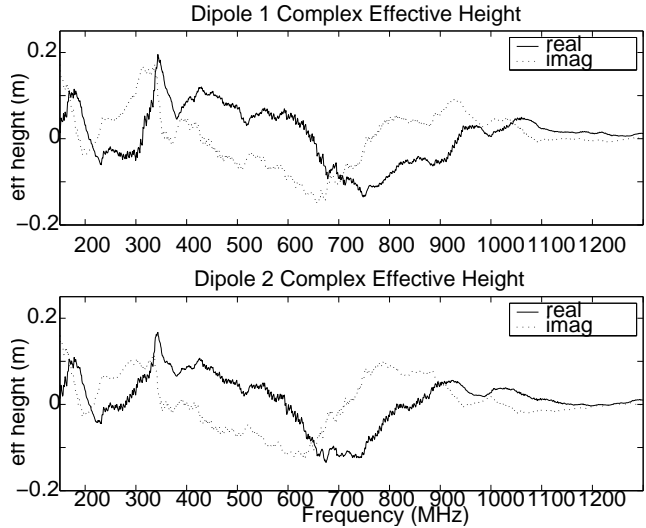}}
\caption{\it Real and Imaginary parts of
RICE complex effective height (in units of meters), 
measured for two dipole antennas in air.}
\label{fig:rice_dipole_H.eps}
\end{figure}

Our current parameterization
of the complex
RICE dipole response is based on time-domain measurements, in air, of
received signals relative to a calibrated standard (Fig. \ref{fig:rice_dipole_H.eps}).
To scale the antenna characteristics in air (complex effective
height ${\vec h_a}$ and complex impedance ${\vec Z_a}$) to ice
(${\vec h_i}'$ and ${\vec Z_i}'$, respectively), we have
used the following procedure: a) assuming that the antenna response
is wavelength-dependent only, we shift the frequency
dependence of the antenna effective height
by the ice index-of-refraction, but assume that the magnitude of the
effective height (both real and imaginary components) remains unchanged
by the new dielectric environment: 
${\vec h_i}'(\omega)={\vec h_a}(n\omega)$ (the variation of the 
peak frequency response of the RICE dipoles
with frequency has been 
qualitatively verified by immersing the dipoles in a large sandbox),
b) the magnitude of the impedance is reduced by $\sqrt{Re(\epsilon)}$; the
frequency dependence is assumed to scale similarly to the effective 
height
${\vec Z_i}'(\omega)={\vec Z_a}(n\omega)/n$ (this scaling
dependence has been verified using ANSOFT FDTD antenna simulations). 
The transfer function is now
re-calculated, by matching the 
scaled antenna impedance to the purely real
$Z_c=50\Omega$
cable load: ${\vec T}'={\vec h_i}'({\vec Z_c}/({\vec Z_c}+{\vec Z_i}'))$.

\subsection{Channel-to-Channel Cross-Talk}
The possibility of spurious hits appearing
in adjacent channels due to cross-talk effects has also been investigated. 
Such effects are observed in
receivers populating the same holes as pulsed transmitters.
Based on the non-observation of time-correlated hits in same-hole receivers 
at the time delays expected from cross-talk, we conclude
that the coaxial cable shielding provides good separation between channels.

\subsection{Hardware Surface Background Rejection \label{ss:HSBR}}
The only recent 
notable modification to the RICE data acquisition (DAQ)
system is the development,
and integration into the DAQ in Jan.
2005, of a Hardware Surface Veto (HSV) board,
designed and developed at the KU Instrumentation Design Laboratory. The
HSV board is a programmable CAMAC module which compares a time-sequence
of antenna hits in a given event trigger with a look-up table of 
time patterns which correspond to anthropogenic surface noise. 
The reference look-up table is updated by the winterover at the South Pole
on a weekly basis.
In the event
of an exact (exclusive) 
match, the trigger is vetoed and the DAQ is reset over the subsequent 
1.2 microseconds. The inefficiency of the HSV board has been checked by
generating simulated surface source events and determining the fraction which
would pass all other neutrino selection criteria. Comparison of the
hit patterns for Monte Carlo simulated
signal events with the patterns used in
our look-up table implies an
inefficiency incurred by the HSV board 
less than 2\%.
 
The utility of the HSV board is assessed numerically by the
winter-over at the South Pole on a weekly basis.
Runs are taken with the HSV board bypassed and with the
HSV board serially inserted into the DAQ chain, in order
to determine the enhancement in livetime with the HSV ON. 
For both HSV-OFF and
HSV-ON runs, we tabulate 
the average livetime (${\cal L}$),
as well as the discriminator threshold (${\cal D}$).
We assume a disk-like sensitive
volume and also that the likelihood of a neutrino at some
source point relative to the detector produces a voltage
$V$ at an antenna, which must exceed ${\cal D}$ in order to produce hits
contributing to the 4-hit trigger, varies inversely with distance,
suggesting the ratio ${\cal L}/{\cal D}^2$ as a measure
of the aggregate neutrino sensitivity. 
Averaged over year 2005 data-taking, the 
estimated gain in sensitivity is approximately 40\%.

\section{Analysis, 
Monte Carlo simulation and Effective Volume\label{sect:MC}}
As described elsewhere\cite{astroph-results}, the RICE Monte Carlo simulation
models the
frequency dependence of the width of the
Cherenkov cone, ice attenuation effects, antenna response, cable 
and amplifier response, and the DAQ electronics. 
Each component in the DAQ chain is characterized 
on the basis of
laboratory, and, where possible, {\it in situ} measurements.
Thermal noise can
be added to each frequency bin assuming
that the thermal power magnitude $P_{thermal}$ 
into the DAQ is given by 
$P_{thermal}=4k(T_{env}+T_{sys})B \sim V_{thermal}/Z^2$,
with $B$ the bandwidth of interest and $T_{env}$ ($\sim$220 K) and
$T_{sys}$ ($\sim$200 K) the environmental and system temperatures,
respectively. The coupling mismatch between the measured antenna
impedance and the purely real 50$\Omega$ DAQ impedance prescribes the
amount of thermal power delivered from the antenna into the DAQ load; within
each frequency bin, the thermal noise amplitude is assigned a random phase
prior to summing with the underlying Cherenkov signal amplitude.
By Fourier transforming this frequency-dependent signal+noise 
spectrum in the last step,
a time-domain signal is produced which can then be compared to the 
measured discriminator response to determine if a trigger signal 
would be generated. Since most neutrinos are at the ``edge'' of 
the detectable volume, the
calculated effective volume with noise is somewhat
larger than the calculated effective volume ignoring noise (more
low-sigma signals passing the trigger threshold after adding 
thermal noise fluctuations than
high-sigma signals failing the trigger threshold after subtracting
thermal noise fluctuations); for the purposes of setting an
upper limit,
the Monte Carlo simulation efficiency results presented
elsewhere in this paper are based on setting the amplitude of thermal noise
to zero. After a simulated event passed this trigger simulation, it
was then embedded into an unbiased event and processed through the
full event reconstruction, yielding the offline software
event detection efficiency $\epsilon$.

The current Monte Carlo simulation improves upon our previous Monte
Carlo in several respects: a) the fully complex transfer function (rather than
only the real portion of the transfer function) is used in determining
the expected signal at the input to the DAQ, b) the current simulation 
uses GEANT4 results on the expected Cherenkov signal strength from 
hadronic showers instead of a 
GEANT3-derived scale factor applied to electromagnetic
showers\cite{addendum}, 
c) geometric distortion of the Cherenkov cone due to variation of
the index of refraction through the firn is included, 
and d) ice dielectric effects are now
based on {\it in situ} measurements recently made at the 
South Pole\cite{RICEnZ,barwickPole}.

\subsection{Signal Shape Modeling \label{ss:SSM}}
Impulsive neutrino signals yield
time-domain responses (``antenna rings'') that 
are largely insensitive to fine details of the neutrino-induced
RF pulse over the RICE frequency bandpass.
Antenna rings
occur provided the time scale of the neutrino signal is much shorter
than the signal
decay time in the radio
receiver. Event-to-event
pulse characteristic variations are largely geometric due to
variations in amplitude and shape across the Cherenkov cone,
as expected in the single-slit source
analogy. Taking two extremes, we have derived the
time-domain signals $V(t)$ from an
Askaryan-like linearly rising 
electric field frequency spectrum $E(\omega)$, and from a
broad-band spectrum, but falling with frequency. These yield
$V(t)$ waveforms nearly indistinguishable in shape.
To assess the possible effect on our offline event reconstruction,
we have examined four different models (``matched filters'') for the
signal shape $V(t)$ expected at the output of the
full DAQ chain: a) channel-by-channel signal shapes based on thermal
noise `hits' observed in the data, b) channel-by-channel signal
shapes based on the data response of each antenna to an
englacial radio transmitter, c) channel-by-channel signal
shapes based on the response of each antenna to a simulated,
short duration neutrino pulse using the 
expected spectral characteristics of
the Askaryan effect, and d) a `general' damped exponential form, which
simply requires that a pulse fit the general profile 
$V(t)=exp(-t/\tau)\cos\omega t$, with $\tau<$30 ns and 
100 MHz$<\omega/2\pi<$500 MHz. 
When applied to a subset of the extant RICE data,
the similarity of these various signal parametrizations in
identifying hits provides confidence in our antenna
calibration and
pattern recognition algorithms.

\subsection{Effective Volume ($V_{eff}$) Calculation \label{ss:VeffCalc}}
The expected 
detected event rate (${\rm GeV^{-1}}$)
can be determined 
using: $N(detected)=V_{eff}\sigma_{\nu N}{\tt n}\Phi\epsilon{\cal L}\Omega$, 
where
$V_{eff}$ is the energy-dependent effective volume
($m^3$), $\sigma_{\nu N}$ is the
neutrino-nucleon cross-section
($m^2$), $\epsilon$ is the software detection
efficiency for an event which is expected to fire the
online hardware trigger ($\sim$0.6), 
{\tt n} is the number density of targets in the
ice ($m^{-3}$), $\Phi$ is the model-dependent flux,
expressed as ($N$/(GeV-$m^2$s-sr)), $\Omega$ is the sensitive solid
angle (sr), and
${\cal L}$ is the livetime. The expected number of 
detected events can then be compared to the observed number of events;
the ratio of these two gives the model-dependent normalization on the flux.

\begin{figure}
\centerline{\includegraphics[width=10cm,angle=0]{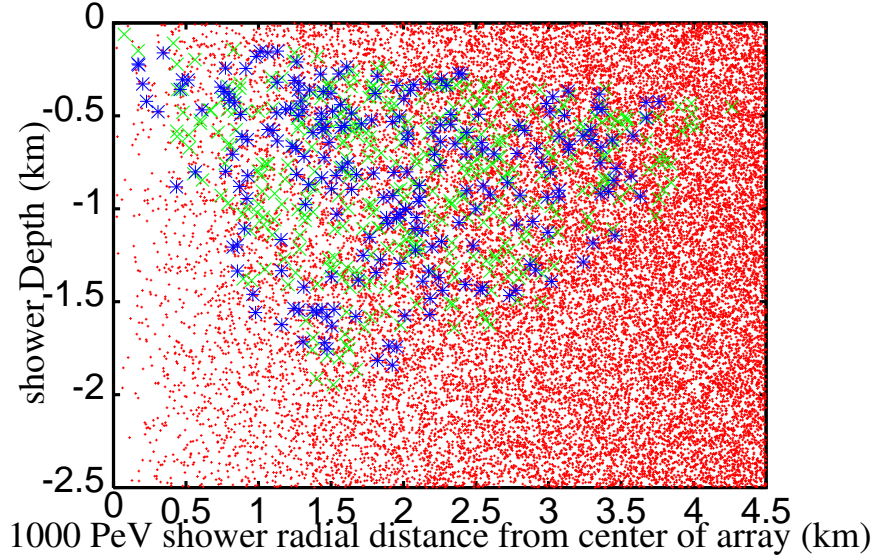}}
\caption{\it Locations of all simulated neutrino interactions (red points) vs.
neutrino interactions producing triggers, for $E_{shower}=1000$ PeV, separately
for electromagnetic (blue asterisks) vs. 
hadronic showers (green crosses), as described in text.}
\label{fig:1000PeV.eps}
\end{figure}

\begin{figure}
\centerline{\includegraphics[width=10cm,angle=0]{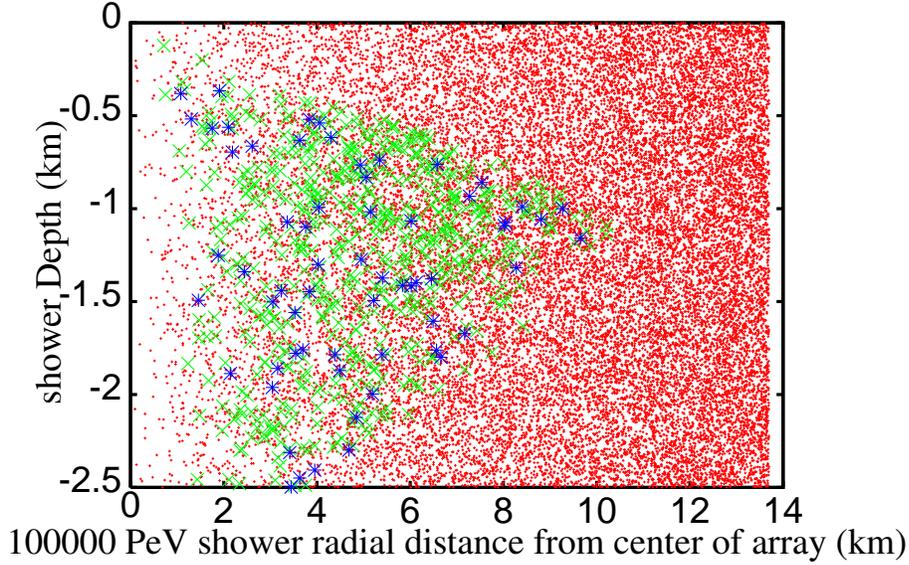}}
\caption{\it Locations of all simulated neutrino interactions (red
points) vs.
neutrino interactions producing triggers, for $E_{shower}=10^5$ PeV, separately
for electromagnetic (blue asterisks) vs. 
hadronic showers (green crosses), as described in text.}
\label{fig:1000000PeV.eps}
\end{figure}

The Monte Carlo effective volume is determined in
two steps. First, 
hadronic and electromagnetic showers are separately
simulated over $2\pi$ sr. Since the earth is nearly opaque at these
energies, flux from the lower hemisphere has only a small effect
on our total effective volume. This contribution to the total effective
volume ($\sim$6\%) is subsequently ignored.
Showers 
initiated by electrons ($e$) are elongated by the LPM effect, whereas 
hadronic ($h$) showers initiated by quark jets are not. As a result 
electron-initiated showers have narrow radiation patterns and exhibit 
reduced detection efficiency.
At a given shower energy, the effective volume is simply calculated
as the ratio of the number of Monte Carlo simulated events which produce
event triggers in the RICE detector, relative to the total number of
simulated events at that energy, multiplied by the total volume sampled:
$V_{eff}(E_{shower})=(N^{MC~triggers}/N^{MC~total~events})*V_0(E_{shower})$.
Since the number of triggers is inversely related to the trigger threshold,
our final experimental result must appropriately sum over the effective
volumes appropriate for the separate running conditions between 1999
and 2005.
Clearly the trigger likelihood
will vary as a function of distance -- at very close distances, 
the possibility of simultaneously
firing four antennas becomes small due to the cone-like
Cherenkov geometry; at large distances, attenuation effects limit the 
efficiency.
Source
distributions of
generated neutrinos (red dots) compared with
reconstructed electromagnetic showers (blue squares) and
reconstructed hadronic showers (green crosses) are shown
in Figures \ref{fig:1000PeV.eps} and \ref{fig:1000000PeV.eps} for two
shower energies. Note that, in the interests of computational execution speed,
we have neglected any possible contribution to the numerator
($N^{MC~triggers}$)
from showers for which the viewing angle,
as measured from the center of the array, differs from the
Cherenkov angle $\theta_c$ by more than 10 degrees ($\approx 1\sigma$ at
f=200 MHz). This biases our result to favor those geometries corresponding to
``direct hits'', with some underestimate of
the total effective volume by ignoring those cases where the
array deviates 
by more than ten degrees from the Cherenkov angle, but the signal
strength is otherwise large enough to result in an event trigger.
As is evident from the Figures, we also neglect any possible
contribution to $V_{eff}$ from the bottom 310 meters of the 
2810-m thick South Polar ice
sheet, given the expected increased 
radiofrequency absorption with temperature\cite{bufordkurtT(z)}.

In the second step, the shower-dependent effective volume is re-cast
as a neutrino-dependent effective volume.
Attenuation and regeneration 
of neutrinos due to earth absorption effects are simulated
separately for $\nu_e$, $\nu_\mu$ and $\nu_\tau$. 
All flavors of neutrino 
create $h$-showers as recoil jets in charged (CC) and neutral (NC) 
current reactions. One flavor ($\nu_e$) creates $e$-showers in CC events. 
For $h$-showers the shower energy is related to the inelasticity
$y(E_\nu)$ and the neutrino energy $E_\nu$ by
$E_s=y(E_\nu)E_\nu$, whereas for 
$e$-showers in $\nu_e$ CC events $E_s=(1-y(E_\nu))E_\nu$.
We use isoscalar-target SM cross sections evolved to 
high energy. The ingredients include the tree-level parton amplitudes and 
CTEQ 6.2 parton distribution functions, with Q$^{2}$ extrapolation where 
required. We also
include a 20\% reduction due to the nuclear (EMC)
effects in 
oxygen. For a given flux model, neutrino mixing is assumed to distribute 
the total flux equally across all three flavors. Due to the competition 
between the LPM effect and an average inelasticity of $y \sim 0.2$, for 
$E_\nu<$1~EeV detection of an isoflavor flux is dominated by 
$e$-showers, whereas $h$-showers dominate above 1 EeV. 


\subsection{Event Reconstruction Efficiency $\epsilon$\label{ss:ERE}}
\begin{figure}
\centerline{\includegraphics[width=15cm,angle=0]{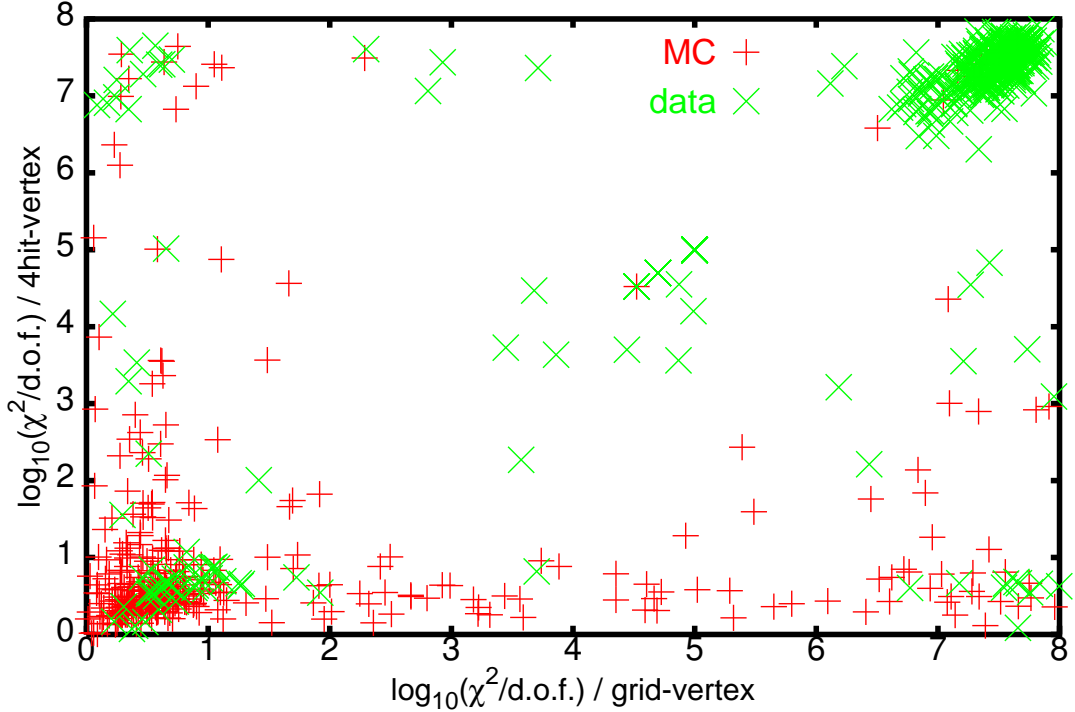}}
\caption{\it Scatter plot of $\chi^2$ for fits
that constrain the
origin of the Cherenkov cone to the vertex determined using
grid-based vertexing algorithm, compared to the vertex found using
the analytic ``4-hit'' vertex algorithm, for events passing
initial time-over-threshold and minimum hit multiplicity 
requirements. Recorded voltages are matched against those
expected for a neutrino-induced shower generating a Cherenkov cone,
with the apex constrained to a given vertex. The cone width is set
to the value expected at the peak of the RICE bandpass ($\sim$300 MHz).
We require that either Cherenkov fit satisfy the
condition $\chi^2/d.o.f.<$50. 
The more distant the reconstructed
vertex, the more difficult it is to discern the curvature of the Cherenkov
wavefront, and the smaller the $\chi^2$; in such a case, all receivers lie in the
same swath of Cherenkov cone.}
\label{fig:chigrid-chi4hit.ps}
\end{figure}

We determine $\epsilon$ by processing
simulated neutrino collision
events embedded into data unbiased events.
We have selected a
background unbiased sample representative of the data comprising
the bulk of our accumulated livetime.
These Monte Carlo events are subsequently analyzed as real data,
and are also tested against the online software veto algorithm.
We note that, although full ray-tracing effects are implemented in neutrino
event
generation, analytic
vertex reconstruction currently ignores ray curvatures and
assumes straight-line trajectories.
Grid-based vertexing properly integrates propagation time over the
full in-ice signal trajectory.
In our simulation, antenna hit
times are smeared by a Gaussian with $\sigma_t$=10 ns (consistent with
timing resolutions derived from transmitter data, but leading to an
overcounting of timing resolution contributions due to hit-recognition
uncertainties, after embedding into unbiased events);
voltages are smeared by $\pm$3 dB to reflect our canonical gain
uncertainty of 6 dB in power.
Figure \ref{fig:chigrid-chi4hit.ps} shows 
a typical simulation vs. data comparison.
Plotted is $log_{10}$ of the $\chi^2$ Cherenkov cone-fit to the hit channels, assuming either the grid vertex or the 4-hit vertex point. 

\begin{figure}[htpb]
\centerline{\includegraphics[width=15cm]{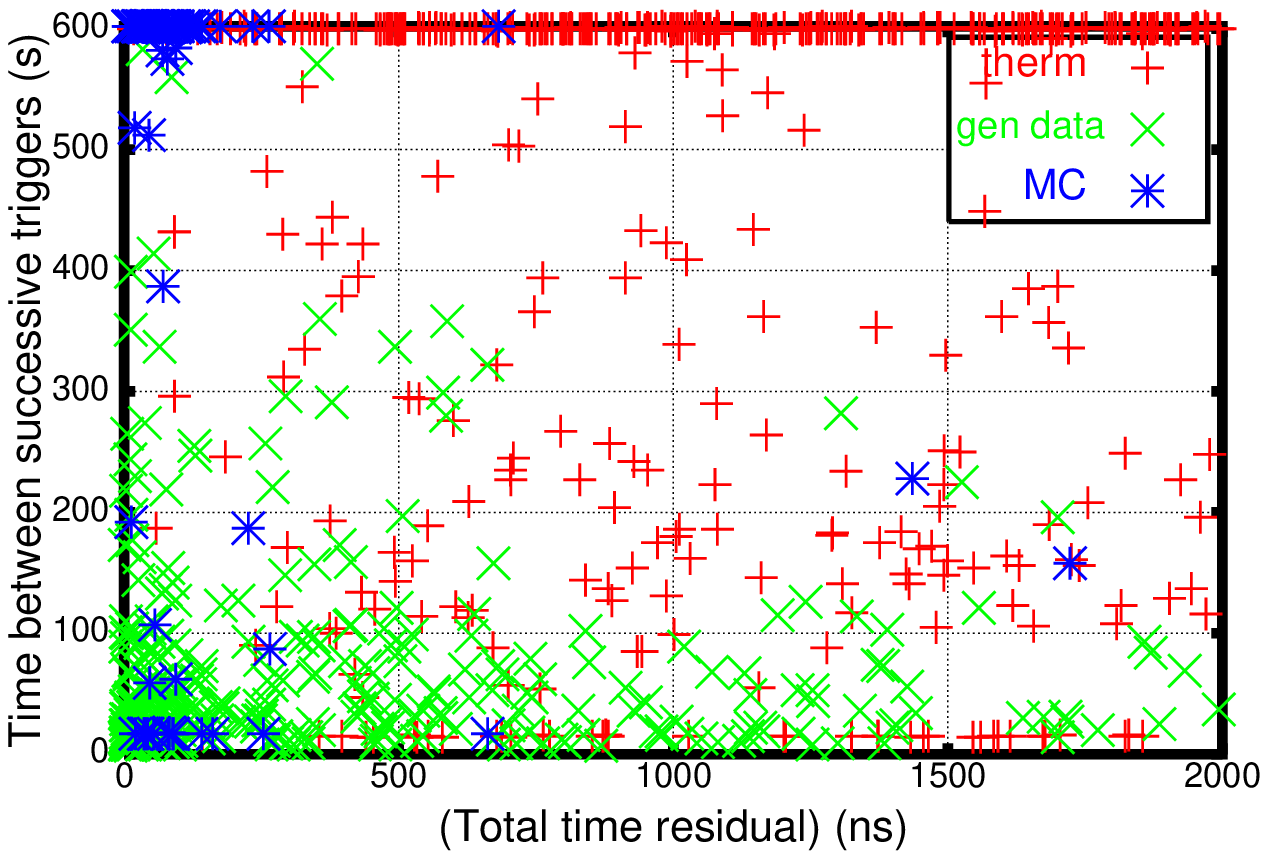}}\caption{\it Time residual vs. time-between-successive triggers for MC neutrino
simulations, general data triggers, and thermal noise fluctuations. In
the absence of an intervening general data trigger, the
10-minute interval between successive forced ``unbiased'' events is evident in the Figure.}\label{fig:thermtrigs}\end{figure}

Similarly, Figure \ref{fig:thermtrigs} compares 
the distribution of the total time residual, summed over all hit channels,
vs. the time-since-last-trigger for
``general''
4-hit triggers (dominated by surface noise),
our signal Monte Carlo sample (embedded into
unbiased events), and a sample of data events which sets a low
threshold (4$\sigma_{rms}$) as a ``hit'' criterion and therefore
preferentially selects thermal noise fluctuations as hits.
As expected, the ``thermal'' sample displays a considerably
broader time residual distribution than the MC neutrino sample.

Once initial offline event selection
requirements (no channels with time-over-threshold greater than
50 ns, and at least four 5$\sigma$ excursions in an event) 
are applied in the first pass,
data are subjected to more rigorous event selection criteria.
This second pass
includes additional cuts, after which
event waveforms are visually examined (hand-scanning).
The effect of the application of the cuts is shown in
Table \ref{tab:cuts};
 ``Data'' refer to typical raw data, 
and ``MC''
gives the survival rate for events 
consisting of
Monte Carlo simulated signal waveforms superimposed upon the
unbiased event `environment'.
The cut values are obtained by comparing Monte Carlo simulated
neutrino events
with events tagged as ``veto'' events by
our fast, online software filter based exclusively on discriminator
threshold-crossing hit times.

\begin{table}[htpb]
\caption{
\it Summary of offline Monte Carlo efficiency and cut application.
(The online software 
surface veto- efficiency is estimated to be $90\pm2$\%, and represents an additional multiplicative inefficiency.)
For the entries below, Monte
Carlo statistical uncertainty is of order 2\% (relative). Systematic
error on reconstruction efficiency is estimated to be $\sim$20\% (relative).
Events passing the last requirement are hand-scanned in the final analysis stage.}
\begin{tabular}{c|c|c|c}
	&	MC (\%) & MC (\%) & Data (\%) \\ \hline
&	EM	& Had	& \\
Selection Requirement &	showers	& shower	& \\ \hline
1) Initial sample       &                     100     & 100       &  100  \\
2) Acceptable Time-Over-Threshold (TOT):      &                     100     & 100    &   39.341 \\
3) $\ge$4 $5|\sigma_{rms}|$ hits:     &                 100  &   100  &   33.223 \\ 
4) $\ge$4 $6|\sigma_{rms}|$ hits:      &                100  &   100  &  16.842 \\
5) Double-Pulse Rejection:  &               99.3  &   99.0 &  16.053 \\
6) High quality 3-d vertex:   &           99.3  &     98.6     &  15.657 \\
7) Vertex depth below firn:    &                 89.9   &   92.8    &   1.119 \\
8) Acceptable Total Time residuals:   &  86.5   &  90.1   &   0.927 \\
9) Passing tighter Time-Over-Threshold:   & 84.0    &   86.0   &   0.919 \\
10) $\le$2 hits with large Time residuals:  & 82.2   &  83.0    &   0.855 \\
11) Acceptable Spatial residuals:      &  81.3     & 79.4   &   0.190 \\
12) Satisfying Cherenkov geometry: & 74.9  & 72.1     &  0.038 \\ 
{\bf 13) $\ge$5 $6|\sigma_{rms}|$ hits:}                   &   67.4    &    66.2    &   0.031 \\ \hline \hline 
\label{tab:cuts}
\end{tabular}
\end{table}

The final hand-scanning stage
removes cases where spurious hits (or incorrectly-determined hit times) 
led to an incorrectly calculated vertex 
location. At this point,
the reconstruction program is fed times
for all channels determined through scanning rather than through
the software pattern-recognition algorithms.
Although the rate of spurious hits is largely antenna-independent,
the vertex displacement relative to the true vertex is obviously antenna
and geometry-dependent. Channel 15, which is roughly twice as deep as the
next-deepest antenna, tends to have disproportionate weight in
calculation of the z-vertex of the event. Many of the events remaining, 
and subsequently discarded, are events for which there was an
incorrect hit-time chosen for channel 15 by the pattern-recognition
algorithm. The presence of `early' hits in the near-surface channels, not used in
vertex reconstruction, but evident in scanning, is also used as a criterion for
rejecting events of surface origin. 
We assume hand-scanning incurs no additional inefficiency in estimating
$\epsilon$.

\begin{figure}
\centerline{\includegraphics[width=10cm,angle=0]{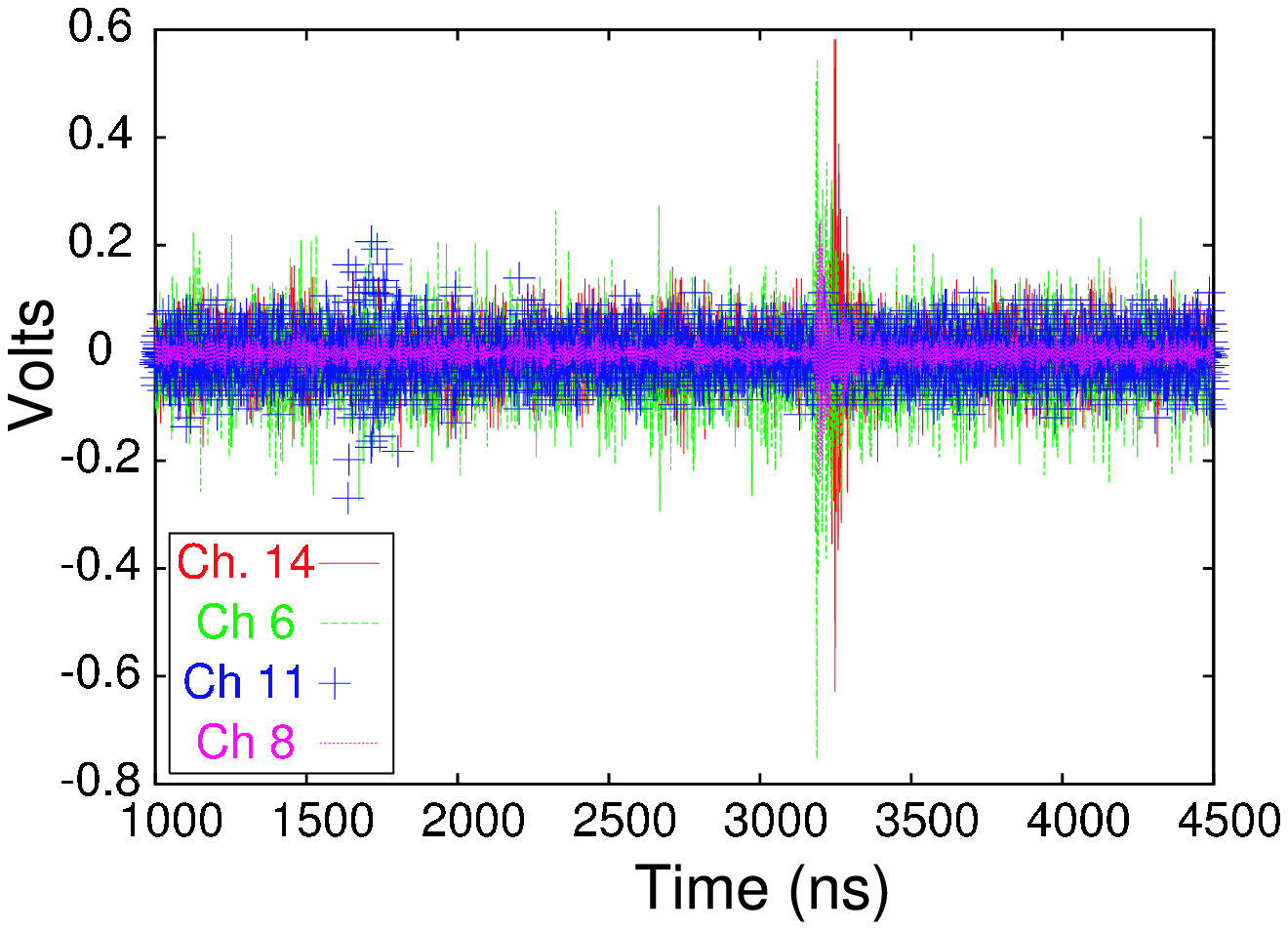}}
\caption{\it Event surviving up to hand-scan, but rejected on the basis
of early hit in channel 11.}
\label{fig:gold-2005.eps}
\end{figure}

No events survive as in-ice shower candidates. An example of an
event which survived initial software requirements, but was
later discarded, is shown in Figure \ref{fig:gold-2005.eps}. For this
event, the reconstructed z-vertex was just below our nominal cut, however,
an examination of the waveforms shows an early hit in channel 11 (points)
at a time approximately $1.6\mu$s prior to hits recorded in 
the deeper channels 6, 8 and 14. Given the
smaller cable delay in channel 11 relative to the other 
channels ($\sim$500-700 ns) and the additional light transit time for 
a source generated at the surface to reach
the deeper channels ($\sim$750-1000 ns), we conclude that
this waveform pattern is consistent with surface-generated backgrounds.

\section{Systematic Uncertainties \label{s:SU}}
Our flux limit is derived directly from the effective volume
$V_{eff}$, the livetime ${\cal L}$, and the event-finding
efficiency $\epsilon(\sim$0.6), which is the product of the
online software veto ($\epsilon_{online}\sim$0.9)
and the offline software veto listed in Table \ref{tab:cuts} 
($\epsilon_{offline}\sim$0.67).
In addition to the aforementioned uncertainty in 
the estimated signal strength produced by a neutrino interaction
($\lesssim$10\%), we have considered several other possible
systematic errors, as detailed below. 

\subsection{Individual Hit Recognition and Event Reconstruction \label{ss:IHR}}
Inaccurate hit-finding will result in events likely to fail our
vertex and time-residual requirements.
To verify our hit-finding algorithms, we have compared the results for
several different hit-definition criteria: a) the maximum voltage
excursion in a waveform, b) the first 6$\sigma$ excursion in a waveform,
c) the time that gives the best match to one
of four `matched filters' (described above).
For simplicity, we
have used as a default option b), although all algorithms give essentially
the same result. Note that the effect of spurious hits is built into the 
inefficiency we quote in our Monte Carlo simulations, which also
include spurious hits in the unbiased events into which simulated showers
are embedded. Once four hits are found, a 3-dimensional vertex is
constructed.
Vertex reconstruction is observed to work well for in-ice sources close
to the array, as calibrated using englacial transmitters.
For sources well outside the array, Monte Carlo
simulations of neutrinos indicate that although directions are generally
well-reconstructed (typical angular deviations between the true and
reconstructed angle to
vertex are of order $\delta(\theta_{true},\theta_{reconstructed})\ll$0.2), 
source depths are generally reconstructed closer to the array than
simulated. This results in an inefficiency in those cases where the true
depth is close to the minimum source depth criterion (200 m).
Events at distances $\gtrsim$500 m (which constitute
the bulk of our sensitive volume) are not
assumed to have well-determined radii, nor are distances to the events
needed for the present analysis.

We assess an overall event reconstruction efficiency
uncertainty of 20\%, based on the limited statistics of 
our Monte Carlo simulation, as well as the variation observed for Monte
Carlo simulations based on different hit-finding
algorithms and using different unbiased event samples.

\subsection{Ray Tracing and Index-of-Refraction\label{ss:R-T}}
\begin{figure}
\centerline{\includegraphics[width=10cm,angle=0]{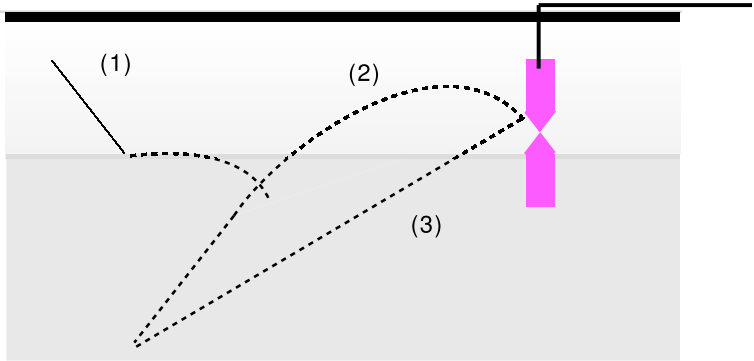}}
\caption{\it Illustration of ray tracing, and possible refractive effects (ray 2), as 
described in text. Dotted
lines indicate possible rays emanating from a neutrino interaction point.}
\label{fig:ray-geometries.eps}
\end{figure}

Since many of our
receivers are located in the firn, radio wavefronts will follow
curved rather than direct-line trajectories, depending on 
the index-of-refraction profile, as discussed
elsewhere\cite{RICEnZ}. This has two significant 
consequences: a) for Cherenkov radiation 
incident at nearly horizontal angles, and angles
slightly below the horizon ($\theta\approx\pi/2$), antennas
in the ``shadow zone'' will not register hits, resulting in
a loss of effective volume. Although initially
directed at the receiver shown, ray 1 in Figure \ref{fig:ray-geometries.eps}
is refracted downwards due to the gradient in the $n(z)$ profile.
b) Due to curvature effects in the firn, a neutrino interaction
below a RICE antenna will, in general, have both a 
``direct'' hit, as well as an ``indirect'' hit (rays 2 and 3 in
the Figure). For the case where ray 3 emerges from the
neutrino interaction point at the
Cherenkov angle, ray 2 emerges with an angle greater than
the Cherenkov angle, with a correspondingly diminished
electric field strength. However, in roughly half the possible
cases, ray 3 will emerge at an angle somewhat smaller 
than $\theta_c$, with ray 2 along $\theta_c$, resulting in a
significant signal from ray 2 due to refractive effects of the
firn. In our Monte Carlo simulation, we now include loss of
effective volume due to shadow-zone effects.
We have not included the expected positive enhancement
in $V_{eff}$ due to ``second-ray''
effects just discussed in our overall systematic error.
An additional possible increase in effective volume is due to the 
focusing of rays, particularly around caustics. This has not
been evaluated numerically and is also not included in our
current calculations.

\begin{figure}
\centerline{\includegraphics[width=10cm,angle=0]{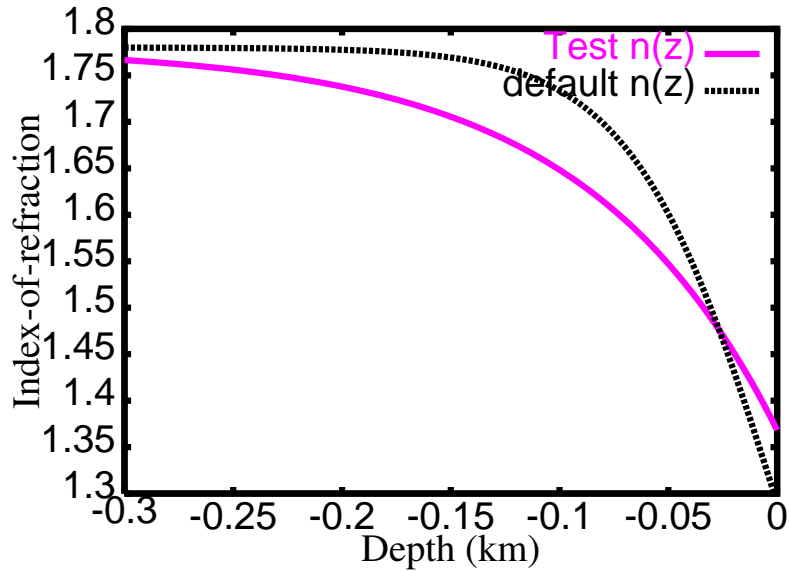}}
\caption{\it Comparison of test profile for index-of-refraction at South Pole with the current RICE default. Former is 
used to assess systematic error due to n(z) uncertainties.}
\label{fig:nfZ-profiles-comp.eps}
\end{figure}

\begin{figure}
\centerline{\includegraphics[width=10cm,angle=0]{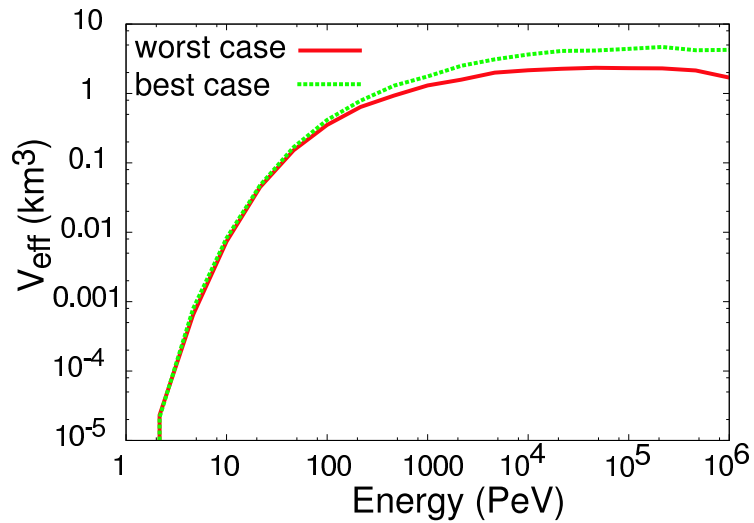}}
\caption{\it Comparison of relative effective volume
(EM showers), obtained without ray tracing 
and ``worst-case'' (``Test n(z)'' in previous Figure) 
variation in index-of-refraction
profile, as a function of depth. Differences are due to fraction
of volume lying in the ``shadow'' region.}
\label{fig:Veffn(z)}
\end{figure}

Neglecting the indirect-hit contributions discussed above,
uncertainties in the real portion of the dielectric constant
are explicitly
evaluated by comparing the effective volumes
using two different models for the index-of-refraction profile
$n(z)$. In Figure \ref{fig:nfZ-profiles-comp.eps},
``Test'' refers to an extreme n(z) profile, inspired by different
measurements of Antarctic ice properties; ``default''
is the profile measured at South Pole\cite{RICEnZ}.
Figure \ref{fig:Veffn(z)} shows the relative $V_{eff}$ obtained using
the test n(z) (``worst case'') vs. $V_{eff}$ obtained without
ray tracing (``best case'').
The effect is largest at high energies where trajectories are 
longest and ray tracing
effects are most important. 

\begin{figure}
\centerline{\includegraphics[width=10cm,angle=0]{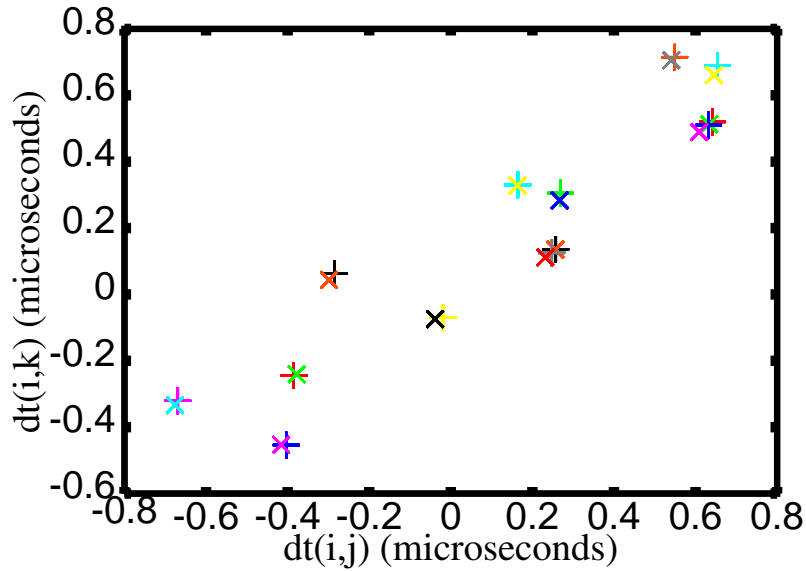}}
\caption{\it Hit Time differences with (``+'') and without (``{\rm x}'')
 ray tracing corrections for channels 0 and 1 (selected at random).}
\label{fig:alltimes.eps}
\end{figure}

For in-ice sources,
the effect of ray tracing corrections on $\epsilon$
is found to be not large, since
vertex reconstruction is based on differences of hit times between
pairs of hit antennas, rather than the absolute transit times from source
location to the
antennas themselves. Figure \ref{fig:alltimes.eps} shows
the time differences between antenna hits
($\delta(t_i-t_j$)) for channels i and j, vs. the same quantity for
channels i and k ($\delta(t_i-t_k)$) for all possible combinations having
$i=0$ or $i=1$. In the Figure, large crosses indicate time differences
obtained without ray tracing, nearby
smaller ``x'' symbols indicate time
differences for the same ensemble of events obtained with ray
tracing. Ray tracing introduces a typical correction of order 5 ns, or
half of our quoted hit time uncertainty.

\subsection{Attenuation Length  \label{ss:ALU}}
As discussed and numerically estimated in our
previous publication, attenuation length uncertainties become
significant at high energies. The corresponding energy dependent 
error in $V_{eff}$ 
is folded into our overall systematic error, using the $\pm1\sigma$ error bars
quoted in the attenuation length measurement made at the South Pole in 2004\cite{barwickPole}.

\subsection{Transfer Function \label{ss:TF}}
\subsubsection{Dipole-to-Dipole Uncertainties \label{sss:DDU}}
These are estimated by comparing the transfer functions measured for
several dipoles. We observe variations in the transfer function of
order 5-10\% in magnitude,
as a function of frequency, resulting in a
relatively small, and symmetric
effect on the calculated neutrino
effective volume.
\subsubsection{Transfer Function scaling from air to ice \label{sss:TFS}}
\begin{figure}
\centerline{\includegraphics[width=10cm,angle=0]{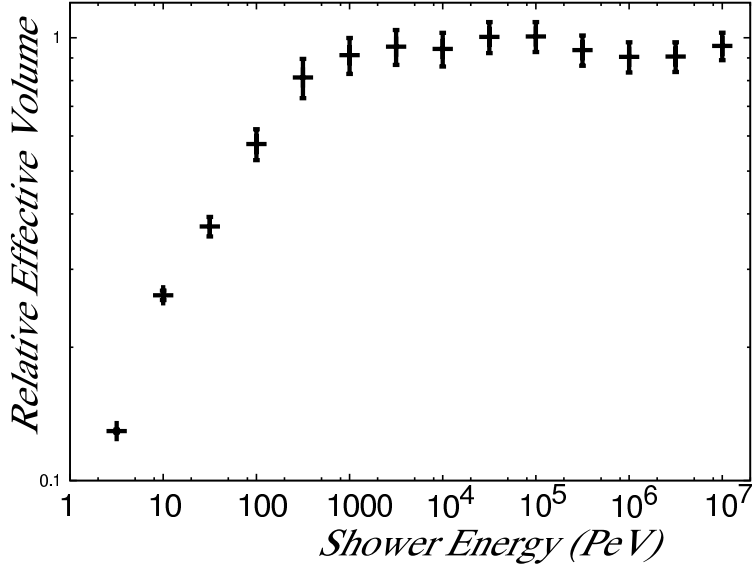}}
\caption{\it Effective Volume dependence on assumed form of transfer function.
Shown is the ratio of effective volumes
(note log y-scale) calculated using an impedance-matching
model for evolving the transfer function in air to ice, 
as done in the current RICE Monte Carlo simulation, relative to a simple
scaling $T_{ice}'(\omega)=nT_{air}(n\omega)$. For E$>$1000 PeV, this ratio
plateaus at a value close to unity.}
\label{fig:Veff-heff.eps}
\end{figure}

To take into account possible uncertainties in our 
scaling from air to ice, we
have compared the effective volume using the scaling
described previously (Section \ref{ss:AR})
to a more extreme scaling where the transfer function is simply
obtained using: $T_{ice}'(\omega)=nT_a(n\omega)$, neglecting the 
complex nature of the transfer function, and assuming that the antenna
is perfectly matched to the cable.
Figure \ref{fig:Veff-heff.eps}
shows the ratio of the effective volumes calculated using these
two different prescriptions. This ratio is folded directly into our
overall relative systematic error. For $E_\nu>$100 PeV, corresponding to
the energy region comprising our greatest effective volume, the effect
is not substantial ($\lesssim$5\%). 
For lower energies, the effect can be considerable,
reflecting the evolution from an $r^3$ sensitive volume to an
$r^2$ volume. 

\subsection{Livetime \label{ss:LT}}
Livetime is calculated online from measurements of the deadtime incurred
per surface veto and also the deadtime incurred per recorded event. We
estimate uncertainties in deadtime to be less than 5\%.
\subsection{Total Gain \label{ss:TG}}
\begin{figure}
\centerline{\includegraphics[width=10cm,angle=0]{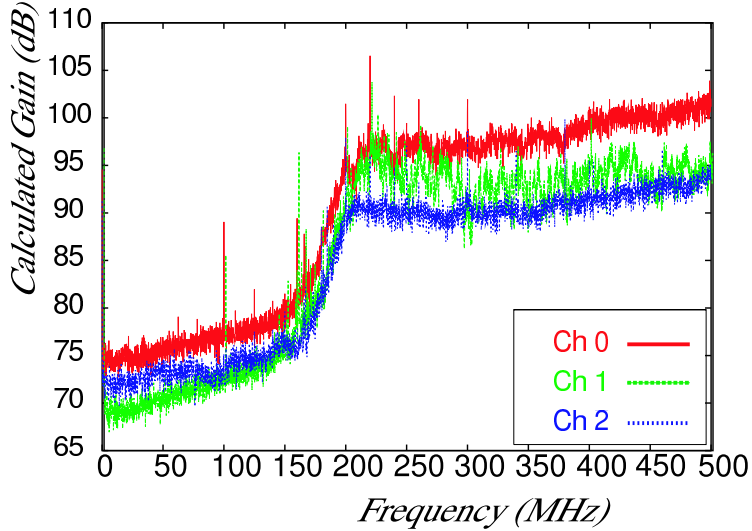}}
\caption{\it Sample channel gains (2002 data).
Smaller calculated gain at low frequencies is
due to explicit filtering of low-frequency backgrounds due to the AMANDA 
experiment, not corrected for in this calculation.}
\label{fig:10-2002.eps}
\end{figure}

The nominal gain uncertainty of $\pm$6 dB in power, although important
at low $E_\nu$, becomes less important at high energies, where ice
absorption effects are dominant. The stability of the
gain of each channel, as a function of frequency, is monitored
online. Figure \ref{fig:10-2002.eps} shows the gains for three sample
channels (2002 data), corrected for cable and insertion losses, but
not low-pass filtering. 
To qualitatively assess the possible implications for our
effective volume, we have run our Monte Carlo code with
our current default amplifier settings
(in the interests of a conservative
upper limit, set to $\sim$90 dB for each channel, slightly below
{\it in situ} calibration in the
RICE bandpass, as shown in Fig. \ref{fig:10-2002.eps}) vs. the August, 2000 
amplifier settings, which were individually tuned, channel-by-channel
to give approximately equal contributions (for each channel) to the
overall discriminator hit rate. To achieve this equality, gains were
turned down in the August, 2000 data sample by as much as 30 dB. 
For large
$E_\nu$, the resulting variation in effective volume is of order 20\%.

\subsection{Birefringence \label{ss:B}}
The possibility of ice birefringence has
been considered\cite{birefs}, 
although there is no evidence to our knowledge for
birefringence of South Polar ice. 
We have not
observed double reception of transmitted pulses or anthroprogenic noise
of mixed polarization\cite{RICEnZ}. 
Non-zero birefringence would lead to an asynchronous 
antenna arrival time of the 
horizontal- vs.
vertical-polarization signal
components ($\sim$25 ns for a source 1 km distant),
resulting in an average loss of signal strength by a factor $1/\sqrt{2}$
for an antenna sensitive to all polarizations. 
Since the RICE dipoles are sensitive to only vertical polarizations,
birefringent effects will not result in an 
expected loss of effective
volume.

\subsection{Final $V_{eff}$ and associated systematic uncertainty 
\label{ss:FV}}
\begin{figure}
\centerline{\includegraphics[width=10cm,angle=0]{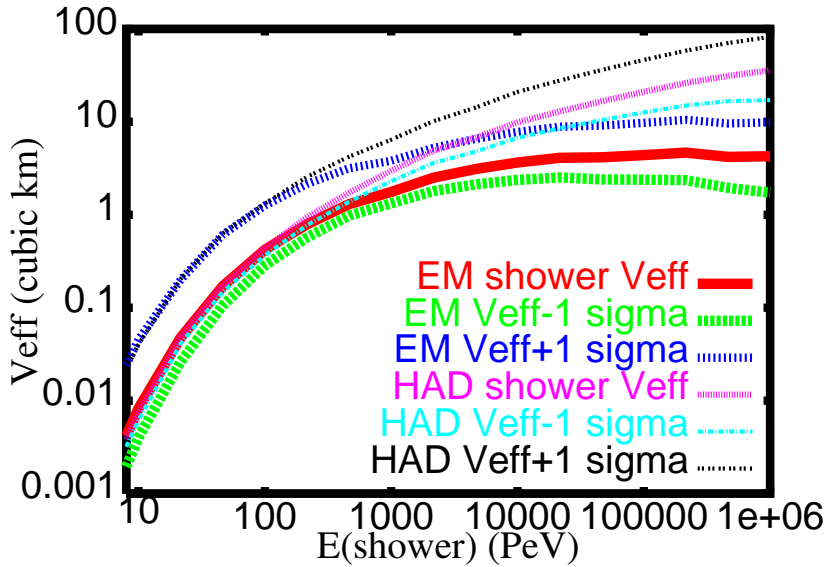}}
\caption{\it Current effective volume for electromagnetic showers
(including LPM effects),
and also hadronic showers, 
with $\pm$1-$\sigma$ systematic errors, as indicated.}
\label{fig:veff.eps}
\end{figure}

Figure
\ref{fig:veff.eps} shows our current effective volume based on
electromagnetic and hadronic 
showers, with $\pm$1-sigma error bars, reflecting the
above systematic errors. At very
high energies, our new central value is approximately a
factor of two smaller than the estimate in our previous publication,
although we again point out that we have purposely excluded 
possible
gains in $V_{eff}$ due to a variety of effects.
Since systematic errors are
not explicitly included in calculation of upper limits, we caution
that our quoted sensitivity has large attendant uncertainties.
Systematic errors
in effective volume, as indicated in Figure \ref{fig:veff.eps} result in
roughly a factor of two possible variation in the 
expected overall neutrino event yield.

\section{Neutrino Flux Limit Results \label{s:NFLR}}

\begin{figure}[htpb]
\includegraphics[width=16cm,angle=0]{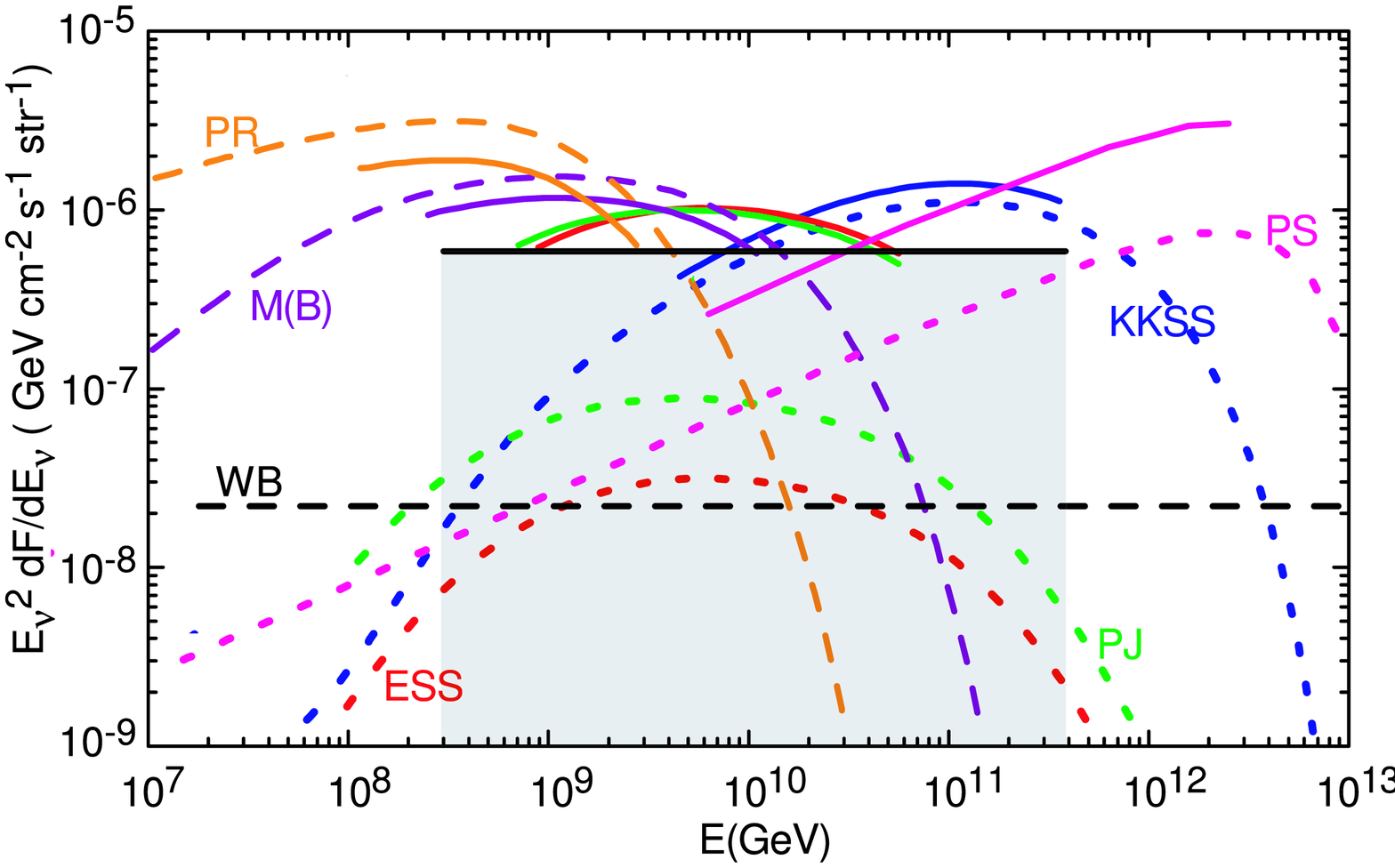}
\includegraphics[width=16cm,angle=0]{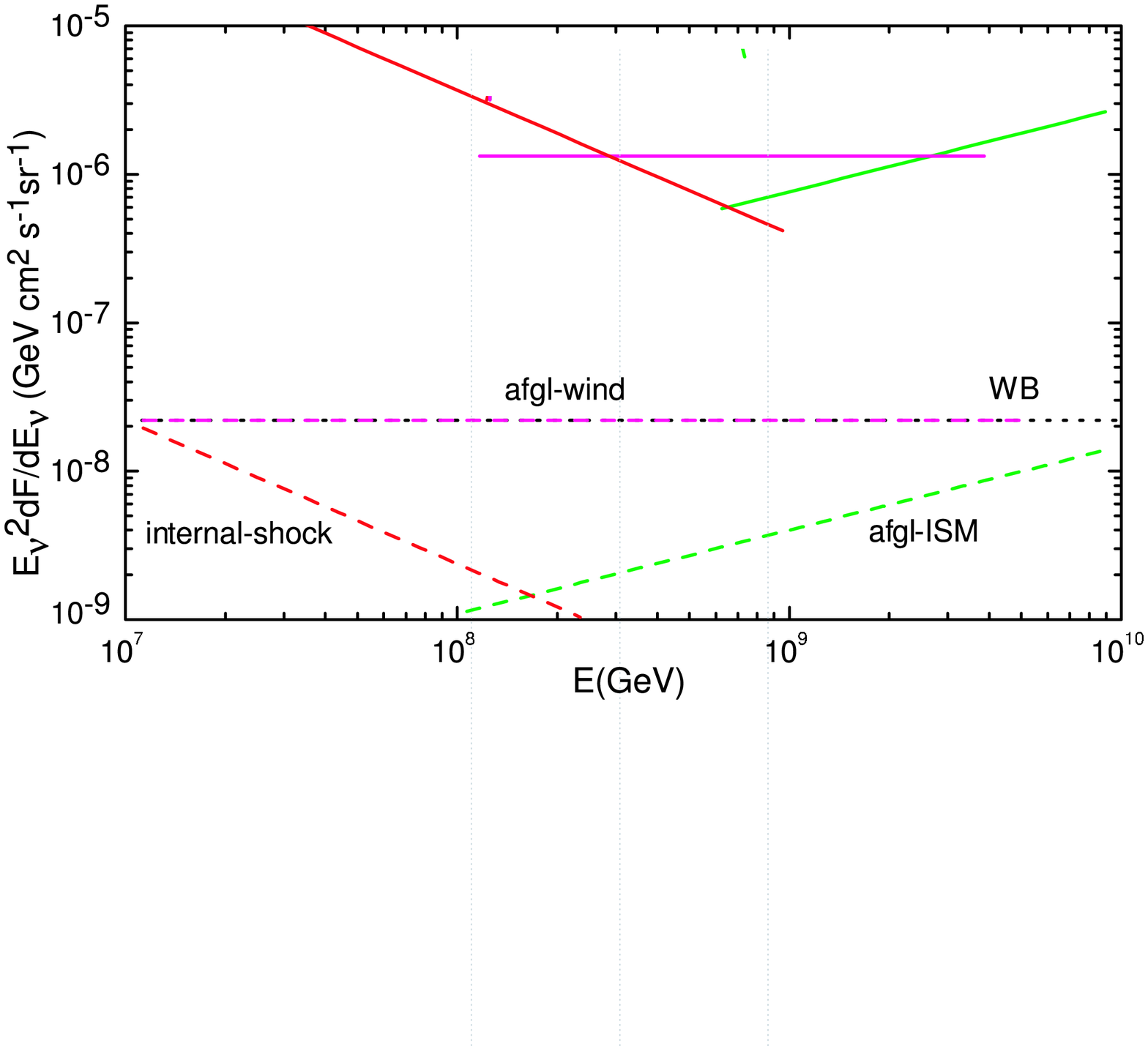}
\vspace{-5.cm}
\caption{\it (Top) Upper bounds on total (all flavor) neutrino fluxes for AGN 
models of PR\cite{pr} 
and MB \cite{mb}, GZK\cite{gzk} 
neutrino models of ESS\cite{ess}, 
PJ\cite{pj}, and KKSS\cite{kkss}, 
and the topological defect model of PS\cite{ps}, due to all flavor 
NC+CC interactions, based on 1999-2005 RICE livetime of about 20500 hrs. 
Dashed curves are for model fluxes and the thick curves are the 
corresponding bounds.  The energy range covered by a bound 
represents the central 80\% of the event rate. (Bottom)
Bounds on diffuse neutrino fluxes from GRBs derived from
RICE data. The bounds are for the
internal shock \cite{wbinsh}, afterglow-ISM \cite{wbism}, and afterglow-wind 
\cite{dailuwind} neutrino flux models assuming an isoflavor
mixture at the detector; we use updated results 
\cite{soebgrbnu} for the fluxes. Systematic errors have not been folded into
calculation of upper limits.}
\label{fig:bnd}
\end{figure}

 
Our 95\% C.L. bounds on 
representative $\nu$-flux models are shown in 
Fig. \ref{fig:bnd}.
The illustrative AGN models are ruled out at 95\% C.L., but the 
Waxman-Bahcall model\cite{wb} is below our limits. The GZK\cite{gzk} flux 
models differ substantially. ESS\cite{ess} and PJ\cite{pj}, 
keyed to models of the stellar
formation rate, are below the RICE sensitivity. The KKSS\cite{kkss} flux, 
constructed to saturate bounds derived from EGRET observations, is just 
barely consistent with our 95\% C.L. limit, i.e. RICE should have 
detected 2 events for this model but observed none. 
Also depicted are 95\% C.L. upper 
limits on diffuse neutrino fluxes predicted
by representative GRB models.

\message{13200 hours???}

\begin{figure}
\centerline{\includegraphics[width=7.5in,angle=0]{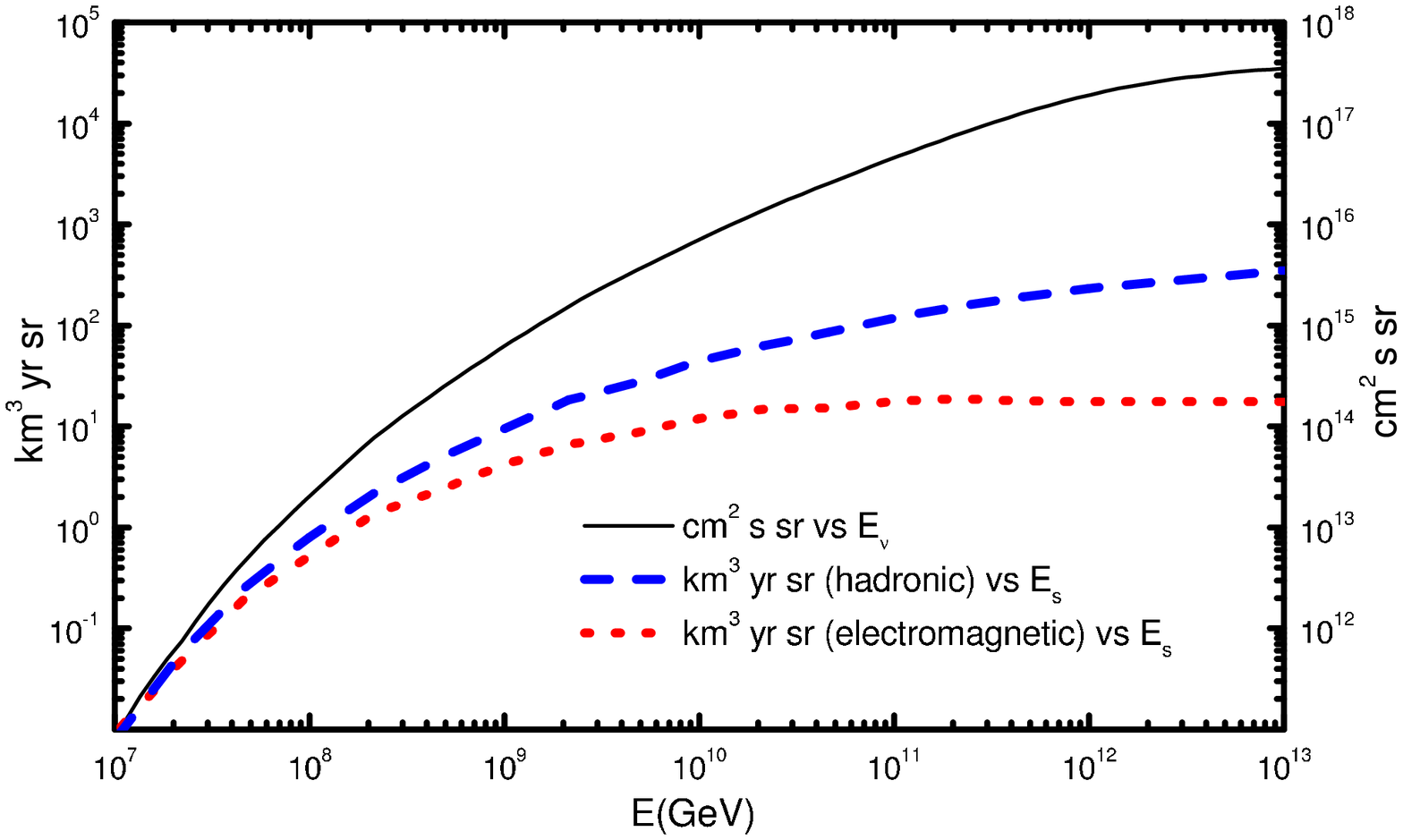}\vspace{-7.5cm}}
\caption{\it Left scale:
Product of $V_{eff}\times Livetime$, as a function of energy. Given
an input flux ($N/(cm^2-sec-sr)$), and a scattering cross-section (and
inelasticity) for any process ($\nu+N\to$electromagnetic shower,
$\nu+N\to$hadronic shower, $\nu+N\to\mu$-black-hole$\to$hadronic shower, 
etc.), one can derive limits on putative cosmic ray fluxes. The 
RICE-specific event reconstruction efficiency has 
already been folded into these curves, and can
therefore be directly convolved with a given input flux to yield
an expected number of events. Right scale: 
Exposure (${\cal A}$); see text for details.}
\label{fig:voltim0005}
\end{figure}

To facilitate application of
our null search to any other possible
related search, Figure \ref{fig:voltim0005} shows the 
livetime-weighted effective volume
(${\cal V}$, with units ${\rm cm^3}$-sr-yr), as a function of energy. 
The y-scale for `hadronic' (dashed) and
`electromagnetic' (dotted) curves is on the left y-axis of the 
Figure; these
two curves show the RICE effective volume integrated over time (1999-2005) and
multiplied by a factor of $2\pi$ steradian, plotted as a function of shower
energy. The difference between `hadronic' and `electromagnetic' curves is
due to the LPM effect. The exposure
${\cal A}$ is indicated by the y-scale for the solid curve (right,
with units cm$^{2}$-s-sr) and
includes standard model NC and CC
cross sections convolved with the effective volume (separately
for hadronic vs.
electromagnetic cases) under the assumption that 1/3 of the total neutrino flux
is $\nu _{e}$. Additional
details on ${\cal V}$
and ${\cal A}$, as well as the procedure
for deriving a predicted RICE observed event yield
given an arbitrary flux model,
are presented in the accompanying Appendix I.

Although the exposure illustrated in Fig. \ref{fig:voltim0005} allows for 
a comparison of models and experiments, it is often desirable to show 
the flux limits from an experiment in a model-independent way, as in Fig. 
\ref{fig:deltalimits}. The procedures used to derive these limits are 
discussed in Appendix II. The bold curve is our best model-independent 
summary of the current RICE results for `typical astrophysics' models. 
The dashed curve represents the envelope of limits for pure power law 
models. The three dotted curves are limits based on logarthmic energy 
bins\cite{AnchordoquiFGS02,GorhamHNLSW04}.

\begin{figure}
\vspace{-4.5cm}
\includegraphics[width=5in,angle=-90]{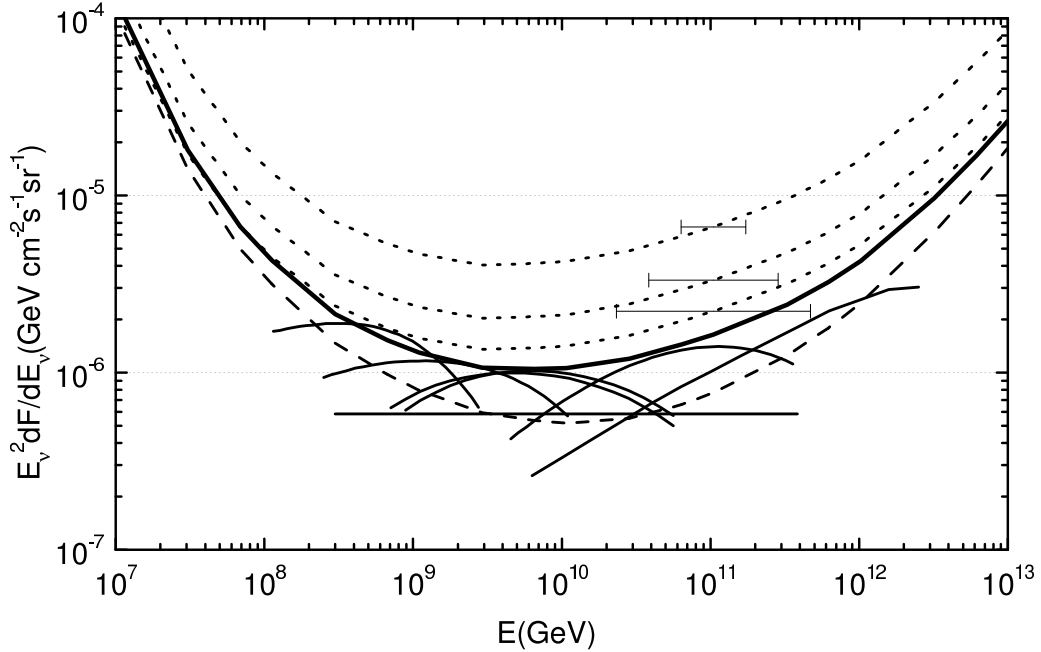}
\caption{\it Model-independent limits for the RICE experiment superimposed on 
model-dependent limits. See Appendix II for discussion.}
\label{fig:deltalimits}
\end{figure}

\section{Summary\label{sect:Summary}}
Using the full dataset accumulated thus far (1999-2005), we 
have presented upper
limits on the incident neutrino flux. 
Despite suboptimal dense-packing of the array, RICE 
provides superior sensitivity in the energy regime between
$10^{17}-10^{20}$ eV. 
Limits are considerably
stronger than previously reported values, and the
most intense flux model projections are ruled out at
95\% confidence level. 

\section{Further Work and Future Plans \label{s:FWFP}}

RICE was originally conceived as a detection system with $km^{3}$ effective
volumes per antenna in the PeV domain\cite{FMR}.  Multi-$km^{3}$ effective volume has been achieved at much higher energies, where fluxes are expected to be much lower. Future
gains in sensitivity will be realized by improvements to several factors now limiting performance: (1) Limited bandwidth of the current experiment arising from cable losses at high frequencies and the 500 MHz bandwidth of the digital oscilloscopes
will be improved with 
high-bandwidth optical fiber 
signal transmission and custom surface digitizer boards, 2) A local
hardware coincidence multiplicity trigger, which only considers ``hits'' for which a local antenna cluster (consisting of 4 antennas) itself satisfies a trigger coincidence
inconsistent with down-coming signals will result in 
enhanced online background rejection.  
As a result, we will greatly improve broad-spectrum energy response by reducing trigger thresholds from the simple one-tier trigger system 
currently in place.  
(3) Improvements in geometric lever-arm and the
detector footprint beyond that of AMANDA/IceCube 
will enhance long-ranged vertex sensitivity. 
During the austral summer of
2005-06, initial deployments of the next generation of neutrino detection hardware are being
made at the South Pole. Details on the hardware itself, as well as the 05-06 deployment, are available from \cite{KUIDL}.

Studies of an expanded radio array are ongoing.
Other technologies that hinge on coherence 
(acoustic detection of showers, e.g.) are also
now being explored by other experimental groups\cite{Learned-acoustic,SAUND,Rolf&Sebastian,Buford-acoustic}, and an {\it in situ} measurement of the acoustic attenuation length at the South Pole is now in progress. 
Preliminary results of the physics potential of a combined radio plus
acoustic detector in conjunction with the IceCube array have recently
been discussed\cite{Rolf&Sebastian}; statistically significant detections of
GZK neutrinos (per year) can be realized at relatively modest costs.

\section{Acknowledgments}
We gratefully acknowledge the generous
logistical support of the AMANDA and SPASE 
Collaborations (without whom this work would
not have been possible), 
the National Science Foundation Office of Polar Programs, the
University of Kansas,
the University of Canterbury Marsden Foundation, 
and the Cottrell 
Research Corporation for their generous financial support. 
Alexey Provorov and Igor Zheleznykh (Moscow Institute of Nuclear
Research, Moscow, Russia) constructed 
the TEM horn antennas currently used as part of the surface-noise
veto. Derek Boyd performed important checks of the overall
system timing calibration.
John Paden and Matt Peters provided essential antenna expertise.
George M. Frichter, Adrienne Juett,
Tim Miller, Dave Schmitz, and Glenn Spiczak all performed
essential work in the initial construction phases of this experiment.
We also thank the winterovers who staffed the
experiment during the last six years at the
South Pole (Xinhua Bai, Allan Baker, Mike Boyce, Phil Broughton, 
Christina Hammock,
Marc Hellwig, Matthias Leupold, Karl Mueller, 
Michael Offenbacher,
Katherine Rawlins,
Steffen Richter and Darryn Schneider),
as well as the excellent on-site support offered by 
Raytheon Polar Services logistical personnel
(particularly Rev. Al Baker, Jack Corbin, Joe Crane and Paul Sullivan).

\vspace{2cm}
\appendix*{\bf
Appendix~I) Calculating upper limits from RICE livetime
${\cal L}$ and $V_{eff}$ 
(Figure \ref{fig:voltim0005})}
\subsection*{Determination of ${\cal V}$}
The `hadronic' (dashed) and
`electromagnetic' (dotted) curves shown in Figure \ref{fig:voltim0005}
are derived as follows: One
computes the effective volume for downward neutrinos using the
standard RICE MC simulation
(accessible from http://kuhep4.phsx.ku.edu/\~~iceman)
separately
for electromagnetic and hadronic showers for different discriminator
thresholds. The output is then integrated over time (weighted by
the amount of data taken at each given 
discriminator setting) to obtain a quantity
with units (km$^3$yr)
as a function of shower energy. This result is
then multiplied by $2\pi$ sr
(for down-coming neutrinos) to obtain a quantity with units
(km$^{3}$yr-sr), defined as ${\cal V}(E_{shower})$. 

\subsection*{Converting from [km$^{3}$yr-sr] to exposure [cm$^{2}$s-sr] 
($\equiv{\cal A}$)}%
We convert ${\cal V}(E_{shower})$ to
${\cal A}(E_\nu)$ as follows:
\[
{\cal A}(E_{\nu })=\epsilon C_{EMC}N_{A}\rho \left[ 
\begin{array}{c}
\int_{y_{0}}^{1}dy\frac{d\sigma ^{NC}}{dy}VT_{hd}(yE_{\nu })+\frac{2}{3}%
\int_{y_{0}}^{1}dy\frac{d\sigma ^{CC}}{dy}VT_{hd}(yE_{\nu }) \\ 
+\frac{1}{3}\int_{0}^{1}dy\frac{d\sigma ^{CC}}{dy}\left( VT_{hd}(yE_{\nu
})+VT_{em}((1-y)E_{\nu })\right) 
\end{array}
\right] ,
\]
where $\epsilon$ is the detector efficiency (0.6 for our case); $C_{EMC}$
(=0.8) is a constant factor used to account for the reduction in
neutrino-nucleon cross sections in oxygen target as opposed to a nucleon
target; $N_{A}\rho$ is Avogadro's number 
multiplied by
the density of ice (0.92 g/cm$^{3}$) which gives total number of target
nucleons per unit volume; $y$ is the inelasticity of the interaction,
and $\frac{d\sigma ^{NC}}{dy}$  
and $\frac{d\sigma ^{CC}}{dy}$ are the neutrino-nucleon
neutral current (NC) and charged current (CC) differential cross sections,
respectively, in the Standard Model. There are three integral
terms. The first term accounts for the contribution from the NC
interactions and is the same for all neutrino flavors; $y_{0}$ is the
lower limit on the integral and is due to the
finite threshold of the detector.
The second term is due to CC interactions of $\nu _{\mu }$ and/or 
$\nu_{\tau }$. The third term is due to the CC interactions of $\nu _{e}$. We
treat $\nu _{e}$ CC interactions separately since both the hadronic
and the leptonic parts of the final products contribute to shower development
in ice; this is not the case for the other two flavors where the lepton
does not contribute to the shower. The factors $\frac{2}{3}$ and $\frac{1}{3}$
are due to the isoflavor
assumption of the model flux, namely,
$\nu_{e}:(\nu _{\mu }+\nu _{\tau })::1:2$.

\subsection*{Significance of ${\cal V}(E_{s})$ and ${\cal A}(E_{\nu })$:}
The quantity ${\cal A}(E_{\nu })$, 
when multiplied with a given model flux d$\phi$/dE$_{\nu }$, 
and then integrated over E$_{\nu }$, gives the expected observed event
rate for 1999-2005 RICE operation. This, in turn, implies bounds on that model
flux under the assumption of the Standard Model 
neutrino-nucleon interactions. The
quantity ${\cal V}(E_{s})$ thus gives one 
freedom to use one's own model for
the neutrino-nucleon interactions and then calculate ${\cal A}(E_{\nu })$ using
the equation above.

\vspace{1cm}
\appendix*{\bf Appendix~II. Model-Independent Neutrino Flux Limits}

\vspace{3ex}

\noindent
The expected number of events observed in an experiment is given by $N = 
\int \phi {\cal A} dE$, where $\phi$ is the flux and ${\cal A}$ is the 
exposure given in units of area $\times$ solid angle $\times$ time. If no 
events are observed, then the 95\% upper limit 
constraint $N<3$ places limits on 
possible flux models. Such model-dependent limits are illustrated in 
Fig. \ref{fig:bnd}. Since it is exhausting to enumerate all models, it is convenient 
to provide a model-independent picture of the strength of an experiment. 
Such model-independent approaches can also be used to compare experiments 
without the bias of choosing a particular model which may favor one 
experiment over another. 

UHE neutrino astrophysics experiments generally have the 
following common properties: a) the exposure increases with energy, b) 
the flux decreases with energy, c) there is almost always a broad 
intermediate energy regime which dominates the event integral $N$. It is 
useful to consider power law models where $\phi \sim E^s$ and ${\cal A} 
\sim E^r$. Then the integral behaves as $N \sim E^t$, where $t= 1+r+s$ 
depends on the combined power laws of flux and exposure. If $t > 0$ then 
$N$ is dominated by high energies, and if $t < 0$ it is dominated by low 
energies. In practice, $t>0$ at low energies due to the increase in 
${\cal A}$, but at higher energies ${\cal A}$ saturates and the flux 
decreases so that $t < 0$. In these circumstances, the event rate is 
dominated by the intermediate energy range around $E_0$, defined by the 
point where $t=0$, or $r+s=-1$. 

One can make an estimate of the event integral by expanding the event 
spectrum around $E_0$. It is convenient to introduce several quantities. 
Define the event spectrum by $g(E) = \phi {\cal A}$. The scaled energy 
and event spectrum are $y = E/E_0$ and $f(y) = g(y E_0)/g(E_0)$. The 
corresponding logarthmic quantities (motivated by the
discussion of power law spectra) are $\eta = \log y$ and $\psi = \log 
f$. 
Accordingly, we define an energy dependent exponent 
$\alpha = d\psi/d\eta = (y/f) df/dy$ and an $E_0$ dependent exponent 
$\beta = \psi/\eta$, i.e. $f = y^\beta$. With these definitions, $N$ can 
be written as 
\begin{eqnarray}
N &=&\int g(E) dE = g_0 E_0 \Delta \\
\Delta &=& \int f(y) dy = \int e^{\psi + \eta} d\eta .
\label{eventcount}
\end{eqnarray}
The $\Delta$ integral is dominated by the region around $E_0$, or 
equivalently the region near $\eta=0$. In this region, $f \sim 1/y$ or 
$e^{\psi + \eta} =1$. Thus, $\Delta$ is the effective range of $\eta$ 
where the event spectrum may be approximated by $1/E$. 

It remains to estimate $\Delta$. Using the definition $\beta=\psi/\eta$ 
and performing a Taylor expansion around $\eta=0$, the integral can be 
recast as $\Delta = \int e^{(1+\beta) \eta} d\eta = \int e^{(1 + \beta_0 
+ \beta'_0 \eta) \eta} d\eta$, where the subscripts denote evaluation at 
$\eta=0$ and the $'$ denotes $d/d\eta$. Using L'Hopital's rule $\beta_0 = 
{{\rm Lim}\above 0pt {\eta\rightarrow 0}} \frac{\psi}{\eta}= \psi' = \alpha(E_0) 
= -1$, and $\beta'_0 = {\rm Lim \above 0pt \eta\rightarrow 0} (\psi' - 
\frac{\psi}{\eta})'/\eta'=\alpha'(E_0) -\beta'_0$, or $\beta'_0=\alpha'_0/2$. Using 
these results in the approximation for $\Delta$, one finds
\begin{equation}
\Delta \simeq \int e^{\frac{\alpha'_0}{2} \eta^2} d\eta = 
\sqrt{\frac{2\pi}{\alpha'_0}} .
\label{delta}
\end{equation}
Once $\alpha'$ and $\Delta$ are determined, the 95\% c.l. 
model-independent flux limit for no observed events is given by
\begin{equation}
\phi_{\rm mi}(E) = 3 \sqrt{\frac{\alpha'}{2\pi}} \frac{1}{E {\cal A}} 
\label{phimi}
\end{equation}

In principle, $\alpha'$ depends on both the exposure and the flux model, 
through the evolution of the exponents $r$ and $s$. For power law flux 
models, or models with weak evolution, we may take $s'=0$. In this case, 
$\alpha'=r'$ may be estimated directly from a log-log plot of the 
exposure. For RICE, we take $r'$ from the exposure shown in 
Fig. \ref{fig:voltim0005}, and find a $\Delta$ which varies from 3 to 9, with 
a peak around $E_0=10^{10.7}$~GeV. The corresponding $\phi_{\rm mi}$ is 
shown as the dashed curve in Fig. \ref{fig:deltalimits}. As is 
apparent, this model-independent limit appears significantly stronger 
than most of the model-dependent limits copied from Figure \ref{fig:bnd}, 
however it is comparable to the limit on the power law $E^{-2}$. (This 
limit is actually a bit weaker than a pure power law, since the event 
integration was cut off at $E_{\rm max}=10^{13}$~GeV.)

It seems clear that neglecting the model evolution $s'$ is not a good 
approximation. To account for this, but still maintain model 
independence, we have recalculated $\alpha'$ taking a constant $s'=0.3$, 
which reduces $\Delta$ to a peak of about 4, but has a lesser effect at 
low energies where $r'$ was larger. The resulting $\phi_{\rm mi}$ is 
shown as the bold solid curve. The middle four `physics' models are well 
fit by this approximation. At low energy, the PR model is evolving faster, 
and the topological model at high energy evolves more slowly, explaining 
the difference from the bold $\phi_{\rm mi}$. 

A similar formalism, where $\Delta$ corresponds to a logarithmic bin 
width, has been used by 
previous authors, but those papers have not focussed on the natural 
choice of $E_0$ defined by the condition $g(E) \sim 1/E$. They have chosen 
$\Delta=1$\cite{AnchordoquiFGS02,GorhamHNLSW04} in an ad hoc manner, or 
$\Delta =3$\cite{ANITA} based on the realization that $\Delta=1$ 
understates model-independent limits relative to model-dependent limits. 
We show model-independent limits with $\Delta = (1,2,3)$ as the three 
dotted curves in the figure. The horizontal error bars graphically show 
the energy range corresponding to those values of $\Delta$.

In summary, for the experimentalist, plots of $\phi_{\rm mi}$ overlaid on 
the same figure serve as a useful method for comparing sensitivities in 
different energy ranges. For the theorist, $\phi_{\rm mi}$ can be 
compared directly to model fluxes. If a model is normalized in such a way 
that the flux is tangent to or intercepts $\phi_{\rm mi}$, then that 
model is ruled out at 95\% c.l. For flux models with evolving spectra, 
an estimate of $s'$ allows for a simple correction to $\phi_{\rm mi}$, 
with a value of $s'=0.3$ being useful for typical GZK models.


\begin{thebibliography}{99}
\bibitem{NEMO}(The NEMO Collaboration),
nemoweb.lns.infn.it/project/htm
\end{thebibliography}
\end{document}